\documentclass[twocolumn,aps,prd,amsmath,amssymb,preprintnumbers,longbibliography]{revtex4-1}
\usepackage{amsmath} \usepackage{amsfonts} \usepackage{amssymb}
\usepackage{bbm}
\usepackage{epsfig}
\usepackage{graphics}
\usepackage{graphicx}
\usepackage{titlesec}
\usepackage{mathtools}
\usepackage{environ}
\usepackage{dsfont}
\usepackage{version}
\usepackage[colorlinks=true]{hyperref}
\hypersetup{urlcolor=black,linkcolor=black,citecolor=black}
\usepackage[toc,page]{appendix}

\makeatletter
\renewcommand*\env@matrix[1][c]{\hskip -\arraycolsep
  \let\@ifnextchar\new@ifnextchar
  \array{*\c@MaxMatrixCols #1}}
\makeatother

\newcommand{\be}{\begin{equation}}
\newcommand{\ee}{\end{equation}}
\newcommand{\ba}{\begin{eqnarray}}
\newcommand{\ea}{\end{eqnarray}}
\newcommand{\nn}{\nonumber}

\titleformat{\subsection}[block]{\normalfont\bfseries}{\thesubsection.}{1ex}{}
\titlespacing{\subsection}{0pt}{10pt}{1pt}[0pt]
\titleformat*{\section}{\large\bfseries}
\renewcommand{\thesubsection}{\arabic{subsection}}

\usepackage{todonotes}
\usepackage{braket}

\definecolor{refkey}{rgb}{0,0,1}
\definecolor{labelkey}{rgb}{0,1,0}

\begin{document}


\title{\LARGE Probabilistic cellular automata\\
for interacting fermionic quantum field theories}



\author{C. Wetterich}

\affiliation{Institut  f\"ur Theoretische Physik\\
Universit\"at Heidelberg\\
Philosophenweg 16, D-69120 Heidelberg}



\begin{abstract}
A classical local cellular automaton can describe an interacting quantum field theory for fermions. We construct a simple classical automaton for a particular version of the Thirring model with imaginary coupling. This interacting fermionic quantum field theory obeys a unitary time evolution and shows all properties of quantum mechanics. Classical cellular automata with probabilistic initial conditions admit a description in the formalism of quantum mechanics. Our model exhibits interesting features as spontaneous symmetry breaking or solitons. The same model can be formulated as a generalized Ising model. This euclidean lattice model can be investigated by standard techniques of statistical physics as Monte Carlo simulations.
Our model is an example how quantum mechanics emerges from classical statistics.
\end{abstract}


\maketitle
\bigskip
\noindent
\section{Introduction}\label{sec: 01}
\medskip

We present in this work a simple classical cellular automaton with homogeneous local updating that describes all features of a unitary interacting quantum field theory with Lorentz symmetry. While the dynamics of the updating is deterministic, the probabilistic aspects of quantum mechanics arise from probabilistic initial conditions for the automaton. The continuum quantum field theory with a continuous unitary time evolution of the wave function emerges as the continuum limit of a discretized theory for infinitely many cells of the automaton and an infinite number of updatings. The continuum limit is also needed for the realization of the continuous Lorentz symmetry. To our knowledge this is the first time that an interacting quantum field theory is described by the deterministic dynamics of a classical system with probabilistic initial conditions. Our simple model may have deep conceptual consequences for our understanding of quantum mechanics. It may also open new methods of computation for fermionic quantum field theories. 

We deal here with classical cellular automata with a local updating, as proposed by Ulam and von Neumann, not with the quantum cellular automata proposed by Feynman, see ref.~\cite{Arrighi2019} for an overview. Our wording "cellular automaton" always refers to the classical local automaton. This type of cellular automaton is used in a wide range of physics and more general science~\cite{WOLF}. We consider a discrete chain of Ising spins with local updating of the configurations within cells of neighboring spins. For "sharp initial conditions" with a well defined initial configuration of Ising spins this is a deterministic system. Such deterministic systems are the basis of t' Hooft's deterministic proposal for quantum mechanics~\cite{GTH}, see also~\cite{ELZE}. 

Probabilistic cellular automata are based on a probability distribution of initial conditions. They constitute simple examples for information transport in classical statistics~\cite{CWIT}. Probabilistic cellular automata correspond to unique jump step evolution operators and assure a "unitary evolution" of the classical wave function, which can be taken as the root of the probability distribution. In a probabilistic view of cellular automata, and more generally of classical probabilistic systems, the concepts of wave functions and density matrices arise in a natural way~\cite{CWPT, CWIT}.
Many features of the quantum formalism are present in classical probabilistic  theories~\cite{CWQF}.
Cellular automata induce a time evolution for which the norm of the wave function is preserved. This property singles out quantum mechanics from more general probabilistic evolution laws. Approximate cellular automata constitute interesting static memory materials~\cite{SEXCW}.

Fermions and Ising spins are closely related, since the binary fermionic occupation numbers $n=(0,1)$ can be directly mapped to Ising spins $s=(-1,1)$. Free fermionic quantum field theories have been realized in this way as generalized Ising models~\cite{CWFCS, CWFGI, CWQFT}. The associated evolution of the quantum wave function corresponds to simple probabilistic cellular automata. There exists a general "fermion-bit map"~\cite{CWFCS, CWFGI} that maps the weight distribution for generalized Ising models to the weight distribution of fermionic models defined by a Grassmann functional integral. This map is based on the construction of the transfer matrix~\cite{BAX, FUC} or the step evolution operator in both formulations, and a proof that the evolution is identical for the generalized Ising model and the associated fermionic theory. The question arises which unitary quantum field theories for fermions can be mapped to a positive weight distribution for Ising spins and vice versa. A positive weight distribution is a classical probability distribution for which many highly developed methods of statistical physics, including efficient numerical simulations, can be employed. Cellular automata are prime examples for a map between probability distributions for Ising spins and unitary fermionic quantum field theories.

So far, generalized Ising models and cellular automata have been constructed only for free fermionic quantum field theories. The present paper provides a first example for an interacting fermionic quantum field theory. We construct explicitly a simple cellular automaton for a particular version of the Thirring model~\cite{THI}. The Thirring model is a rather simple two-dimensional model that allows for exact solutions~\cite{THI, KLA, FUR, NAO, AAR, DNS, FAIV}. Nevertheless, it admits rich features as spontaneous symmetry breaking and solitons, which  have been used by Coleman~\cite{COL} for a map to a bosonic model. We will see all these features directly in the evolution of our classical automaton.

The cellular automaton constructed in the present paper  corresponds to a particular model in this class, characterized by a special value of an imaginary coupling. Despite the coupling being imaginary, the quantum field theory is unitary, as implied directly by its equivalence to a cellular automaton. Our model can indeed describe spontaneous symmetry breaking and solitons. It can be investigated either by a numerical or analytic solution for the cellular automaton, or by methods of fermionic quantum field theories. We also construct the associated generalized Ising model. This allows for the use of Monte-Carlo simulations or similar numerical techniques. It may provide for a new direction how fermionic quantum field theories could become accessible to numerical simulations.

We start in sect.~\ref{sec: 02} by introducing the quantum formalism for the description of classical cellular automata. This introduces the wave function and the evolution operator for a conceptually simple description how the probabilistic information of the initial condition is processed as "time" progresses for consecutive steps in the automaton. This formalism applies to the classical  automaton and does not constitute a new model. The key concept for our purpose is the step evolution operator that encodes the updating law of the automaton. We introduce in this section the local homogeneous updating rule for our automaton, representing a particular type of Thirring model. 

In sect.~\ref{sec: 03} we present the general construction of the step evolution operator for fermionic quantum field theories which are formulated as a Grassmann functional integral. We discuss a first simple model for which the step evolution operator is identical to the one of a cellular automaton. In this very simple example the map between wave functions and their evolution in a quantum system and the corresponding objects for probabilistic cellular automata becomes apparent. We also proceed to the continuum limit. 

In sect.~\ref{sec: 04} we introduce two-dimensional fermionic quantum field theories for complex Dirac spinors. We specify the particular fermion model that is equivalent to the cellular automaton in sect.~\ref{sec: 02}. The interaction corresponds to a particular type of Thirring model. Lorentz symmetry emerges in the continuum limit. We discuss in sect.~\ref{sec: 05} how many characteristic features of the fermionic quantum field theory as spontaneous symmetry breaking or solitons can be seen directly in the evolution of the cellular automaton. 

In sect.~\ref{sec: 06} we briefly discuss that our particular Thirring type model can also be formulated as a generalized Ising model. This is an euclidean functional integral on a two-dimensional lattice. Sect.~\ref{sec: 07} discusses there findings, stressing the emergence of quantum mechanics from classical statistics \citep{Wetterich:2020kqi}.

\bigskip
\noindent
\section{Probabilistic cellular automata}\label{sec: 02}
\medskip

Cellular automata can be described by a series of discrete time steps $t$, $t+\varepsilon$, $t+2\varepsilon$ etc., starting at some initial time $t_\textup{in}$. Every state $\rho$ at $t$ is transformed to a state $\overline{\tau}(\rho)$ at $t+\varepsilon$, and so forth for increasing time.
To be specific, we consider at each $t$ a number $M$ of Ising spins  $s_{\gamma}(t)$, $\gamma=1\dots M$, $s^{2}_{\gamma}(t)=1$.
The $N=2^{\scriptstyle M}$ states $\rho$ are the possible configurations of Ising spins. A given cellular automaton is characterized by a sequence of maps $\rho \rightarrow \overline{\tau}(\rho)$ for a sequence of time steps. These maps can be represented by $N\times N$ - matrices. We consider here invertible cellular automata for which at every $t$ the inverse map $\tau\rightarrow\overline{\rho}(\tau)$ exists. 

\bigskip
\noindent
{\bf Step evolution operator and wave function for \\ cellular automata}
\medskip

A convenient formalism uses for each time step the step evolution operator $\widehat{S}_{\tau\rho}(t)$, given by 
\begin{equation}\label{eq: A01}
\widehat{S}_{\tau\rho}(t)=\delta_{\tau, \overline{\tau}(\rho)}=\delta_{\overline{\rho}(\tau), \rho} \;.
\end{equation}
It is a matrix with precisely one element equal to one in each row and column, and zeros otherwise. The state of the system at $t$ can be described by a vector or "wave function" $q(t)$ with $N$ components $q_{\rho}(t)$. 
The evolution law for a cellular automaton reads
\begin{equation}\label{eq: A02}
q_{\tau}(t+\varepsilon)=\widehat{S}_{\tau\rho}(t)q_{\rho}(t)\; .
\end{equation}
Consider first a deterministic cellular automaton which has at each time step a well defined configuration of Ising spins~$\sigma$. For a given state $\sigma$ at $t$ one takes $q_{\rho}(t)=\delta_{\rho,\sigma}$, such that only one component of the vector $q$ differs from zero. This implies
\begin{equation}\label{eq: A03}
q_{\tau}(t+\varepsilon)=\delta_{\overline{\rho}(\tau),\rho} \,\delta_{\rho, \sigma}=\delta_{\overline{\rho}(\tau),\sigma}=\delta_{\tau, \overline{\tau}(\sigma)}\; , 
\end{equation}
and zero otherwise. The state $\sigma$ at $t$ is indeed transported to $\overline{\tau}(\sigma)$ at $t+\varepsilon$. The sequence of time steps corresponds to an ordered matrix multiplication of step evolution operators
\begin{equation}\label{eq: A04}
q_{\tau}(t+n\varepsilon)=\biggl[\widehat{S}\big(t+(n-1)\varepsilon\bigr) \:\dots\: \widehat{S}(t+\varepsilon)\widehat{S}(t)\biggr]_{\tau\rho} \!\!q_{\rho}(t)\;.
\end{equation}

The step evolution operator encodes the updating rule of the automaton. For most automata one considers "homogeneity in time" with the same $\widehat{S}(t)$ for all $t$, repeating always the same updating. We will consider here two alternating updatings $\widehat{S}(t)$ for even $t$ and $\widehat{S}(t+\varepsilon)$ for odd $t$, repeated sequentially. One may consider the product $\widehat{S}(t+\varepsilon)\widehat{S}(t)=\overline{S}(t)$ as a combined updating rule for steps from $t$ to $t+2\varepsilon$. For the combined updating our automaton is homogeneous in time.

Probabilistic cellular automata are cellular automata for which the initial condition is given by a probability distribution $\big{\lbrace} p_{\tau}(t_{\textup{in}})\big{\rbrace} , \, p_{\tau}\geqslant 0,\, \sum_{\tau}p_{\tau}=1$. It is convenient to associate at every $t$ the probability distribution to a "classical wave function"~\cite{CWQPCS, CWPT, CWFCS} 
\begin{equation}\label{eq: A05}
p_{\tau}(t)=q^{2}_{\tau}(t)\; .
\end{equation}
Probabilistic initial conditions are then encoded in a more general form of the initial wave function $q(t_{\textup{in}})$, given by a unit vector with
\begin{equation}\label{eq: A06}
p_{\tau}(t_{\textup{in}})=q^{2}_{\tau}(t_{\textup{in}}) , \quad \sum_{\tau}q^{2}_{\tau}(t_{\textup{in}})=1 \;.
\end{equation}

Indeed, the map of states $\rho\rightarrow\overline{\tau}(\rho)$ translates to a map $q_{\overline{\tau}(\rho)}(t+\varepsilon)=q_{\rho}(t)$, $p_{\overline{\tau}(\rho)}(t+\varepsilon)=p_{\rho}(\tau)$, as realized by eq.~\eqref{eq: A02}. In consequence, the probability distribution at $t>t_{\textup{in}}$ is obtained from eqs.~\eqref{eq: A04}, ~\eqref{eq: A05}. The relation $p_{\tau}(t)=q^{2}_{\tau}(t)$ guarantees the positivity and normalization of the probability distribution for every $t>t_{\textup{in}}$, since $\widehat{S}$ is an orthogonal matrix which preserves the norm of $q$. Expectation values of local observables at every $t$ can be extracted from the probability distribution $\lbrace p_{\tau}(t)\rbrace$ by the standard rules of classical statistics.

\bigskip
\noindent
{\bf Alternating switch gates}
\medskip

A simple example is the "switch operator" for two bits, $M=2$. We label the four states $\tau = 1 \dots 4$ by the configurations of Ising spins $(1,1)\,,\, (1,-1)\,,\, (-1, 1)\,,\, (-1,-1)$. For the analogy with fermions we actually will rather use occupation numbers $n_{\gamma}=(s_{\gamma}+1)/2$, $n_{\gamma}=1,0$ \;, and label the states by $(1,1)\,,\, (1,0)\,,\, (0,1)\,,\,(0,0)\;$. The switch operation maps
\begin{align}\label{eq: A07}
S: &(1,0) \,\leftrightarrow\,(0,1)\,,\notag \\
&(0,0)\,,\,(1,1)  \;\textup{invariant}. 
\end{align}
The corresponding step evolution operator reads
\begin{equation}\label{eq: A08}
\widehat{S}_{S}= \begin{pmatrix}
1 & 0 & 0 & 0 \\ 
0 & 0 & 1 & 0 \\ 
0 & 1 & 0 & 0 \\ 
0 & 0 & 0 & 1
\end{pmatrix}  ,\; \widehat{S}^{2}_{S}=1\;. 
\end{equation}
This is a rather trivial automaton for which two particles of different type are interchanged at every time step.

Cellular automata describing Thirring type models can be based on an alternating sequence of generalized switch operations or switch gates.
We start with the simplest type of switch gates.
At a given $t$ the spins are on a one-dimensional chain with positions $x$ separated by $\varepsilon$, e.g.  $s_{\gamma}(t)=s(t, x)$.
Correspondingly, at every $t$ the states are distributions of fermionic particles, given by occupation numbers $n(t, x)$.
We define even and odd $t$ or $x$ by $t=m_{t}\varepsilon ,\; x=m_{x}\varepsilon$~, with even or odd integers $m_{t}, m_{x}$.
At even $t$ we consider pairs of neighboring positions $x$ and $x+\varepsilon$, with $x$ even. They define blocks with two sites. The cellular automaton moves a single particle at $x$ to $x+\varepsilon$, and vice versa. 
Configurations with two particles or zero particles in the block remain invariant. If we associate the two species of particles with particles at $x$ and $x+\varepsilon$, this amounts to the switch~\eqref{eq: A07},~\eqref{eq: A08}. The switch operation acts independently on all pairs or in all blocks. 

At $t+\varepsilon$ we change the grouping into pairs of neighbors at $x$ and $x-\varepsilon$, for even $x$. The switch operates now between the particles in the new blocks. As a result, $\widehat{S}(t)$ and $\widehat{S}(t+\varepsilon)$ do not commute. In the following, we alternate the pairing, $\widehat{S}(t+2\varepsilon)=\widehat{S}(t)$. Repetition of a sequence of two switches does no longer produce unity as in eq.~\eqref{eq: A08}, since
\begin{equation}\label{eq: A09}
\bigl(\widehat{S}(t+\varepsilon)\widehat{S}(t)\bigr)^{2}=1+\widehat{S}(t+\varepsilon)\widehat{S}(t)\bigl[\widehat{S}(t+\varepsilon), \widehat{S}(t)\bigr]\; .
\end{equation}

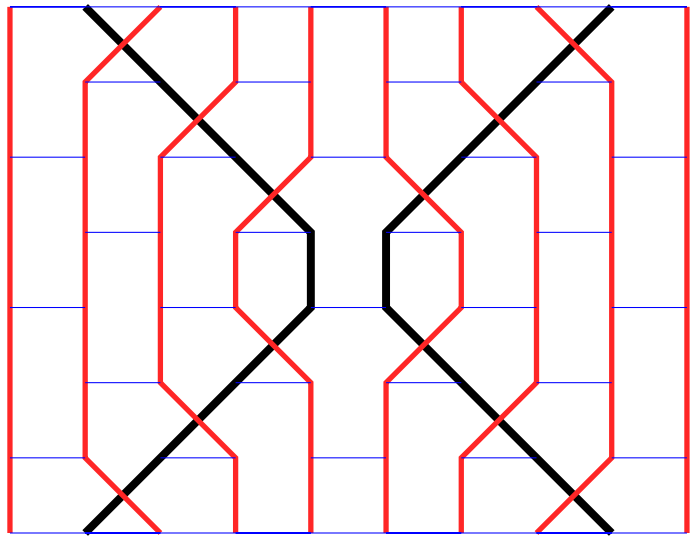
\begin{figure}[h]
\begin{tikzpicture}
\draw[black, line width=3pt] (1,0)--(4,3)--(4,4)--(1,7)
			(8,0)--(5,3)--(5,4)--(8,7);
\draw [red!85, line width=2pt](0,0) -- (0,7)
			(2,0)--(1,1)--(1,6)--(2,7)
			(3,0)--(3,1)--(2,2)--(2,5)--(3,6)--(3,7)
			(4,0)--(4,2)--(3,3)--(3,4)--(4,5)--(4,7)
			(5,0)--(5,2)--(6,3)--(6,4)--(5,5)--(5,7)
			(6,0)--(6,1)--(7,2)--(7,5)--(6,6)--(6,7)
			(7,0)--(8,1)--(8,6)--(7,7)
			(9,0)--(9,7);
\draw [blue](0,0)--(9,0)
			(0,7)--(9,7)
			(0,1)--(1,1)
			++(1,0)--++(1,0)
			++(1,0)--++(1,0)
			++(1,0)--++(1,0)
			++(1,0)--++(1,0)
			(1,2)--(2,2)
			++(1,0)--++(1,0)++(1,0)--++(1,0)++(1,0)--++(1,0)++(1,0)
			(0,3)--(1,3)++(1,0)--++(1,0)++(1,0)--++(1,0)++(1,0)--++(1,0)++(1,0)--++(1,0)
			(1,4)--(2,4)++(1,0)--++(1,0)++(1,0)--++(1,0)++(1,0)--++(1,0)
			(0,5)--(1,5)++(1,0)--++(1,0)++(1,0)--++(1,0)++(1,0)--++(1,0)++(1,0)--++(1,0)
			(1,6)--(2,6)++(1,0)--++(1,0)++(1,0)--++(1,0)++(1,0)--++(1,0)
			(0,7)--(1,7)++(1,0)--++(1,0)++(1,0)--++(1,0)++(1,0)--++(1,0)++(1,0)--++(1,0)
			(1,0)--(2,0)++(1,0)--++(1,0)++(1,0)--++(1,0)++(1,0)--++(1,0);
\end{tikzpicture}
\caption{Particle scattering. The black lines link occupied sites, the red lines empty sites. We have indicated the pairing for the switch gates by thin blue lines.}\label{Fig.1}
\end{figure}

As a result, this rather simple automaton allows already for a non-trivial evolution. For only a few particles present, and most positions empty, the evolution can be viewed as the propagation of the particles on light cones with occasional scattering, as depicted in Fig.~\ref{Fig.1}.
The totally empty and the totally occupied states are invariant. For the regular half-filled state with alternating occupied and empty sites one finds the same state at $t+2\varepsilon$ and $t$.
For this state all particles are either freely right moving, and empty sites or holes left moving, or all particles are left moving with right moving holes. These features show qualitatively already many analogies to fermions.

\bigskip
\noindent
{\bf Local updating}
\medskip

We will see below that this automaton describes actually free fermions. In Fig.~\ref{Fig.1} we could equivalently draw the black lines as crossing without scattering. Interacting fermions can be obtained if one takes two different species or colors of particles at every site, say red and green. The updating proceeds again in blocks of two sites, with an alternating choice of pairs as above. For each block one has to update 16 states, corresponding to the 16 configurations of four different occupation numbers. The updating rules are as follows:
\begin{enumerate}
\item\ If the lower left corner is occupied by a single red particle (R) and the lower right corner by a single green particle (I), the red particle moves to the upper left corner and the green particle to the upper right corner. They jump one unit in $t$ without changing position $x$. The same rule holds if red and green colors are interchanged.\label{item: 1}
\item If the lower left corner and the lower right corner are both occupied by a single red particle, the upper left corner and upper right corner are both occupied by a single green particle. Both particles move on as $t$ progresses, but change color. The same holds if the red and green colors are interchanged. \label{item: 2}
\item For all other configurations except the ones of ~\ref{item: 1}.)~and~\ref{item: 2}.) the occupation numbers switch site as $t$ increases, from lower left to upper right and lower right to upper left. \label{item: 3}
\item The blocks alternate, comprising at $t$ the sites $x$ and $x+\varepsilon$, and at $t+\varepsilon$ the sites $x$ and $x-\varepsilon$, where $t$ and $x$ are taken even.\label{item: 4}
\end{enumerate}

For a simple illustration we can again take the scattering picture of Fig.~\ref{Fig.1}. If one of the black lines for occupied states is a red particle, and the other one green, the lines continue after scattering with the same color. If the two incoming particles are red, the two outgoing particles are green, and vice versa.

\begin{figure}[h]
\noindent
\resizebox{20pc}{!}{%
\begin{tikzpicture}
\draw[red!85, line width=2pt] (0,0) -- (4.5,4.5) -- (6.5,2.5) -- (4,0) 
(0,6) -- (1.5,7.5) -- (3.5,5.5) --(5.5,7.5) -- (3.5,9.5) -- (4,10)
(10,10) -- (6.5,6.5) -- (8.5,4.5) -- (13.5,9.5) -- (13,10)
(12,0) -- (12.5,0.5) -- (10.5,2.5) -- (14,6)
(8,0) -- (8.5,0.5) -- (9,0);
\draw[black!60!green, line width=3pt](0,9) -- (1.5,7.5) -- ( 3.5,9.5) -- (3,10)
(0,2) -- (3.5,5.5) -- (4.5,4.5) -- (6.5,6.5) -- (5.5,7.5) -- (8,10)
(8.5,0.5) -- (6.5,2.5) -- (8.5,4.5) -- (10.5,2.5) -- (8.5,0.5)
(13,0) -- (12.5,0.5) -- (14,2)
(14,9) -- (13.5,9.5) -- (14,10);
\draw [blue] (1,7) rectangle (2,8);
\draw [blue] (3,9) rectangle (4,10);
\draw [blue] (3,5) rectangle (4,6);
\draw [blue] (4,4) rectangle (5,5);
\draw [blue] (5,7) rectangle (6,8);
\draw [blue] (6,6) rectangle (7,7);
\draw [blue] (6,2) rectangle (7,3);
\draw [blue] (8,4) rectangle (9,5);
\draw [blue] (8,0) rectangle (9,1);
\draw [blue] (10,2) rectangle (11,3);
\draw [blue] (12,0) rectangle (13,1);
\draw [blue] (13,9) rectangle (14,10);
\draw[dashed, blue]	(11,1)rectangle(12,2);
\draw[dashed, blue]	(9,1)rectangle(10,2);
\draw[dashed, blue]	(12,2)rectangle(13,3);
\draw[dashed, blue]	(10,0)--(10,1);
\draw[dashed, blue]	(11,0)--(11,1);
\draw[dashed, blue]	(13,1)--(14,1);
\draw[dashed, blue]	(13,2)--(14,2);
\draw[dashed, blue]	(13,1)--(13,2);
\draw[dashed, blue]	(10,1)--(11,1); 
\draw [blue] (9.5,1.5)  circle (1mm);
\draw [blue] (10.5,0.5)  circle (1mm);
\draw [blue] (11.5,1.5)  circle (1mm);
\draw [blue] (12.5,2.5)  circle (1mm);
\draw [blue] (13.5,1.5)  circle (1mm);
\draw [blue] (-0.2,9.4) -- (0,9.7)--(0.2,9.4);
\draw [blue] (13.4,-0.2) -- (13.7,0)--(13.4,0.2);
\draw [blue] (0,0) -- (13.5,0);
\draw [blue] (13.5,0) -- (14,0) 
node[below, font=\huge] at (13.7,-0.2){$x$};
\draw [blue] (0,0) -- (0,9.5);
\draw [blue] (0,9.5) -- (0,10)
node[left, font=\huge] at (-0.2, 9.7) {$t$};
\draw [blue] (0,10) -- (14,10);
\draw [blue] (14,0) -- (14,10);
\draw [blue] (-0.1,0) -- (0,0)
++(-0.1,1) --++(0.1,0)
++(-0.1,1) --++(0.1,0)
++(-0.1,1) --++(0.1,0)
++(-0.1,1) --++(0.1,0)
++(-0.1,1) --++(0.1,0)
++(-0.1,1) --++(0.1,0)
++(-0.1,1) --++(0.1,0)
++(-0.1,1) --++(0.1,0)
++(-0.1,1) --++(0.1,0);
\draw [blue] (14,0) -- (14.1,0)
++(-0.1,1) --++(0.1,0)
++(-0.1,1) --++(0.1,0)
++(-0.1,1) --++(0.1,0)
++(-0.1,1) --++(0.1,0)
++(-0.1,1) --++(0.1,0)
++(-0.1,1) --++(0.1,0)
++(-0.1,1) --++(0.1,0)
++(-0.1,1) --++(0.1,0)
++(-0.1,1) --++(0.1,0);
\draw [blue](0,0) -- (0,-0.1)
++(1,0.1) --++(0,-0.1)
++(1,0.1) --++(0,-0.1)
++(1,0.1) --++(0,-0.1)
++(1,0.1) --++(0,-0.1)
++(1,0.1) --++(0,-0.1)
++(1,0.1) --++(0,-0.1)
++(1,0.1) --++(0,-0.1)
++(1,0.1) --++(0,-0.1)
++(1,0.1) --++(0,-0.1)
++(1,0.1) --++(0,-0.1)
++(1,0.1) --++(0,-0.1)
++(1,0.1) --++(0,-0.1)
++(1,0.1) --++(0,-0.1)
++(1,0.1) --++(0,-0.1);
\draw [blue](0,10.1) -- (0,10)
++(1,0.1) --++(0,-0.1)
++(1,0.1) --++(0,-0.1)
++(1,0.1) --++(0,-0.1)
++(1,0.1) --++(0,-0.1)
++(1,0.1) --++(0,-0.1)
++(1,0.1) --++(0,-0.1)
++(1,0.1) --++(0,-0.1)
++(1,0.1) --++(0,-0.1)
++(1,0.1) --++(0,-0.1)
++(1,0.1) --++(0,-0.1)
++(1,0.1) --++(0,-0.1)
++(1,0.1) --++(0,-0.1)
++(1,0.1) --++(0,-0.1)
++(1,0.1) --++(0,-0.1);
\end{tikzpicture}}
\caption{Cellular automaton for a particular Thirring model.
 We show single particle lines for red and green particles. Blocks for which interactions take place are indicated by solid squares. In the lower right part we also indicate by dashed boundaries and marked by a small circle a few blocks in which no interaction takes place. We do not indicate additional two-particle lines where a red and a green particle are on the same site and move together on diagonals without scattering.}\label{Fig.2}
\end{figure}
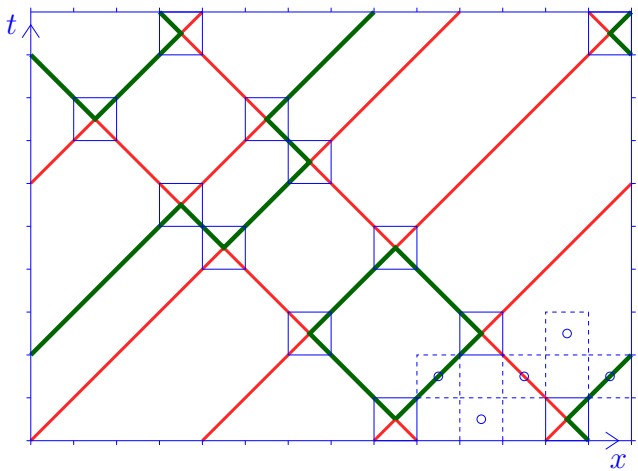

We have depicted the evolution of a configuration with a few particles in Fig.~\ref{Fig.2}. 
\noindent
Single particles move on diagonal straight lines. They change color whenever they meet another single particle. In addition, there are two particle lines (not shown in Fig.~\ref{Fig.2}) where a red and a green particle move together on the same diagonal trajectory without scattering. Fig.~\ref{Fig.2} also demonstrates that the cellular automaton is invariant under $\pi/2$-rotations in the (t,x)-plane. Zooming on the two scatterings in the third row from below we observe scattering events as the one shown in Fig.~\ref{Fig.1}, now for two particles with different color which cannot be identified. Note that the drawing of colored lines inside the scattering boxes is a particular picture. We also could draw the lines on the vertical boundaries of the box, as in Fig.~\ref{Fig.1}. In Fig.~\ref{Fig.2} we have not indicated lines for empty sites (or holes). For two holes on each site we could draw a cross, or vertical lines as in Fig.~\ref{Fig.1}.

For initial conditions with a unique sharp configuration at $t_{\textup{in}}$ the dynamics is solved by registering the crossings of single particle lines and accounting for the change of colors. It is obvious that all non-trivial dynamics is related to color and cannot be detected by color-blind observables. A fixed configuration of occupation numbers of fermions with sharp values of one or zero is a possible initial state for the particular Thirring-type model. It is an allowed quantum state since the occupation number operators at different sites $x$ commute. Only for this particular initial condition the quantum wave function evolves according to a deterministic cellular automaton.
For smooth wave functions the automaton is probabilistic.

Pairs of green particles can be created or annihilated, there is a loop of green particles, or a green particle line can move "backwards in time". These are typical features of the path integral in a quantum field theory. We will show that this automaton indeed describes a full fledged quantum field theory for interacting fermions.

The updating rule is a local rule. For a single time step the state of each block at $t+\varepsilon$ involves only the occupation numbers within the same block at $t$. We may combine two time steps to a composite time step from $t$ to $t+2\varepsilon$ in order to realize homogeneity in time. At $t+2\varepsilon$ the states within a block at $x$ and $x-\varepsilon$, for which the switches occur from $t+\varepsilon$ to $t+2\varepsilon$, involve at $t$ the occupation numbers at $x-2\varepsilon$, $x-\varepsilon$, $x$, $x+\varepsilon$.
One observes a causality structure with a past lightcone. Only occupation numbers within this past lightcone can influence the outcome at a given $t$.

\bigskip
\noindent
\section{Step evolution operator for fermions}\label{sec: 03}
\medskip

We next develop the language of a fermionic quantum field theory for this and similar cellular automata. It is based on fermion-bit map~\cite{CWFCS, CWFGI} between Ising spins and fermions in the occupation number basis.
The central idea is to construct the step evolution operator for the fermionic theory and to establish that it is identical to the step evolution operator of the cellular automaton. The step evolution operator in a quantum field theory acts on a multi-particle wave function. Usually, this wave function is complex. We can write it in a real formulation with the double number of components, consisting of the real and imaginary parts of the complex wave function. For an identical step evolution operator the quantum wave function obeys then the same evolution law \eqref{eq: A04} as the classical wave function for the cellular automaton. Also the rule for the computation of expectation values of local observables in terms of the wave function $q(t)$ is the same for the fermionic quantum field theory and the cellular automaton. There is a one-to-one map between the two formulations - they describe the same physics.

\bigskip
\noindent
{\bf Grassmann functional integral}
\medskip

In order to establish an isomorphic map between some discretized version of a type of Thirring model for fermions and the particular cellular automaton described above we extract the step evolution operator from the weight function for fermions in a formulation with Grassmann variables~\cite{CWFCS, CWFGI}. 
The microscopic fermionic quantum field theory is formulated as a Grassmann functional integral with an action $S[\psi]$. The action is a functional of the Grassmann variables $\psi_{\alpha}(t,x)$. We investigate first a discrete version with a finite number of sites $(t,x)$ and a finite number of variables $\psi_{\alpha}(t,x)$. The action is then a function of the Grassmann variables. The continuum limit will be taken subsequently.

We assume that the weight function $w_{0}[\psi]$ can be written as a product of local factors $\tilde{\mathcal{K}}(t)$
\begin{equation}\label{eq: A10}
w_{0}[\psi]=\exp\big{\lbrace}-S[\psi]\big{\rbrace}=\prod_{t}\tilde{\mathcal{K}}(t) \;,
\end{equation}
where $S[\psi]$ is the action and $\tilde{\mathcal{K}}(t)$ depends on Grassmann variables on neighboring $t$-layers $\psi_{\gamma}(t)$ and  $\psi_{\gamma}(t+\varepsilon)$. Here $\gamma$ is a collective index including space-labels, $\gamma=(x,\alpha)$.
Each local factor $\tilde{\mathcal{K}}(t)$ can be written as a polynomial in the Grassmann variables involved. We assume that all terms involve an even number of Grassmann variables, such that local factors at different $t$ commute.

An element of a local Grassmann algebra is a function of the local Grassmann variables $\psi_{\gamma}(t)$, where locality refers here only to $t$. It is a linear combination of basis elements $g_{\tau}$, 
\begin{equation}\label{eq: A11}
\tilde{q}(t)=q_{\tau}(t)\, g_{\tau}\bigl[\psi(t)\bigr] \;.
\end{equation} 
Each basis element is a polynomial of factors $\psi_{\gamma}(t)$, where we take some ordering with the smaller $\gamma$ to the left, with a conveniently chosen sign.
Since at each "place" $\gamma$ in the polynomial there can be either a factor $\psi_{\gamma}$ or one (i.e.~no~$\psi_{\gamma}$), there are $N=2^{\scriptscriptstyle M}$ possibilities, similar to the configurations of $M$ occupation numbers. We identify a factor $\psi_{\gamma}$ with an empty site, and a factor one with an occupied site. An example with $M=2$ is 
\begin{align}\label{eq: A12}
(1,1):\quad g_{1}&=\phantom{-} g_{1}'=\phantom{-} \overline{g}_{4}=\phantom{-} \overline{g}_{4}'=1 \,, \nn \\
(1,0): \quad g_{2}&=\phantom{-} g_{2}'=-\overline{g}_{3}=-\overline{g}_{3}'=\psi_{2}  \, , \nn \\
(0,1): \quad g_{3}&=\phantom{-} g_{3}'=\phantom{-} \overline{g}_{2}=\phantom{-} \overline{g}_{2}'=\psi_{1} \, , \nn \\
(0,0): \quad g_{4}&=-g_{4}'=\phantom{-} \overline{g}_{1}=-\overline{g}_{1}'=\psi_{1}\,\psi_{2} \; . 
\end{align}
We will later identify the coefficients $q_{\tau}(t)$ for $t$ even with the wave function of the probabilistic cellular automaton, with local probabilities $p_{\tau}(t)=q^{2}_{\tau}(t)$. Our aim is to find the evolution law which permits the computation of $\left\lbrace q_{\tau}(t+\varepsilon)\right\rbrace$ from $\left\lbrace q_{\tau}(t)\right\rbrace$.

We also introduce conjugate basis elements $\overline{g}_{\tau}$ by the relation 
\begin{equation}\label{eq: A13}
\int\! \mathcal{D}\, \psi \,\overline{g}_{\tau}[\psi]\,g_{\rho}\,[\psi]=\delta_{\tau\rho} \;.
\end{equation}
Here all elements are local at $t$, and the local Grassmann integral is given by $\int\!d\psi_{\scriptscriptstyle M}(t) \dots d\psi_{1}(t)$ .
In order to organize the signs arising from the anticommutation of Grassmann variables in an efficient way, we also employ
\begin{equation}\label{eq: A14}
g'_{\tau}=(-1)^{\frac {\scriptstyle m_{\tau}(m_{\tau}-1)}{2}} g_{\tau}\quad , \quad \overline{g}_{\tau}'= (-1)^{\frac{\scriptstyle \overline{m}_{\tau}(\overline{m}_{\tau}-1)}{ 2}} \overline{g}_{\tau}\; ,
\end{equation}
with $m_{\tau}$ the number of $\psi$ factors in $g_{\tau}$, and $\overline{m}_{\tau}$ the number of $\psi$\,-\,factors in $g_{\tau}$ . The different sets of basis functions obey
\begin{equation}\label{eq: A15}
\exp(\psi_{\gamma}\varphi_{\gamma})=g_{\tau}(\psi)g_{\tau}'(\varphi)=\overline{g}_{\tau}'(\psi)\overline{g}_{\tau}(\varphi)\; ,
\end{equation}
and
\begin{equation}\label{eq: A16}
\!\!\int\!  \mathcal{D}\psi g_{\tau}'(\psi) \overline{g}_{\rho}' (\psi)= \eta_{\scriptscriptstyle M}\delta_{\tau\rho}\quad ,\quad \eta_{\scriptscriptstyle M}=(-1)^{\frac{\scriptstyle M(M-1)}{2}}\; .
\end{equation}

\bigskip
\noindent
{\bf Step evolution operator from Grassmann \\ functional integral}
\medskip

A Grassmann functional integral defines a sequence of step evolution operators or normalized transfer matrices. For this purpose we expand for even $t$ the local factor as 
\begin{equation}\label{eq: A17}
\tilde{\mathcal{K}}(t)=\overline{g}_{\tau}'(t+\varepsilon) \widehat{S}_{\tau\rho}(t)\,\overline{g}_{\rho}(t) \; .
\end{equation}
The coefficients $\widehat{S}_{\tau\rho}(t)$ will be identified with the matrix elements of the step evolution operator. With eq.~\eqref{eq: A13} one finds
\begin{equation}\label{eq: A18}
\begin{split}
\int\!  \mathcal{D}\psi(t)\,\tilde{\mathcal{K}}(t)\,\tilde{q}(t)=\widehat{S}_{\tau\rho}(t)\,q_{\rho}(t)\,\overline{g}_{\tau}'(t+\varepsilon) \\
=q_{\tau}(t+\varepsilon)\,\overline{g}_{\tau}'(t+\varepsilon)\; .
\end{split}
\end{equation}
Here we expand for odd $t$
\begin{equation}\label{eq: A19}
\tilde{q}(t)=q_{\tau}(t)\,\overline{g}_{\tau}'\bigl[\psi(t)\bigr] \;,
\end{equation}
instead of eq.~\eqref{eq: A11} for even $t$. According to the modulo two properties in time we employ for odd $t$ the expansion
\begin{equation}\label{eq: A20}
\tilde{\mathcal{K}}(t)=g_{\tau}(t+\varepsilon)\, \widehat{S}_{\tau\rho}(t)\,g_{\rho}'(t)\; .
\end{equation}

Integrating out the variables at $t+\varepsilon$ the product of local factors results in matrix multiplication ($t$ even)
\begin{equation}\label{eq: A21}
\begin{split}
\int\!  \mathcal{D}\psi &(t+\varepsilon)\tilde{\mathcal{K}}(t+\varepsilon)\tilde{\mathcal{K}}(t) \\
&=\eta_{\scriptscriptstyle M}g_{\tau}(t+2\varepsilon)\bigl[\widehat{S}(t+\varepsilon)\widehat{S}(t)\bigr]_{\tau\!\rho} \overline{g}_{\rho}(t) \,.
\end{split}
\end{equation}
Considering longer chains of neighboring local factors and integrating out intermediate Grassmanns variables one arrives at eq.~\eqref{eq: A04}, multiplied on both sides with appropriate Grassmann basis functions.
(A possible factor $(-1)$ from powers of $\eta_{\scriptscriptstyle M}$ will be omitted - if necessary it can be absorbed by a slight redefinition of expansions.) In general, the matrices $\widehat{S}$ defined by eqs.~\eqref{eq: A17},~\eqref{eq: A20} are the transfer matrices. We normalize $\tilde{\mathcal{K}}(t)$ by multiplication with a constant such that the largest eigenvalue of $\widehat{S}$ obeys $\left| \lambda\right|=1$. With this normalization the transfer matrix becomes the step evolution operator.

We will construct fermionic models for which the step evolution operator $\widehat{S}$ contains in each row and column precisely one non-zero element. With a suitable normalization and sign convention of the Grassmann elements these elements can be chosen to equal one. We obtain in this way the unique jump operator $\widehat{S}$ of a cellular automaton. We identify the wave function in the fermionic model with the wave function of the probabilistic cellular automaton. This "fermion-bit map" allows us to describe the evolution of the local probability distribution for cellular automata by a fermion model that produces the same step evolution operator. The fermion-bit map can be extended to maps of observables, not necessarily involving only variables at a given time~\cite{CWFGI}.
Our aim will be to find fermionic models which represent a given cellular automaton, as the one described above.

\bigskip
\noindent
{\bf Simple fermion model}
\medskip

A first Thirring type model is defined by a Grassmann functional integral
\begin{equation}\label{eq: 01}
Z=\int\!  \mathcal{D}\psi \exp\lbrace-S[\psi]\rbrace=\int\!  \mathcal{D}\psi \exp \lbrace-i S_{\scriptscriptstyle M}[\psi]\rbrace \; ,
\end{equation}
with euclidean action $S$ related to the action $S_{\scriptscriptstyle M}$ for Minkowski signature by $S=iS_{\scriptscriptstyle M}$. The action is local and invariant under Lorentz-transformations, $S=\int_{t}\! \mathcal{L}(t)$, with
\begin{equation}\label{eq: 02}
\mathcal{L}(t)=-\int_x\! \big{\lbrace}\overline{\psi}\gamma^{\mu}\partial_{\mu}\psi-\frac{g}{2}(\overline{\psi}\gamma^{\mu}\psi)(\overline{\psi}\gamma_{\mu}\psi)\big{\rbrace} \;.
\end{equation} 
With two-dimensional Dirac matrices given by the Pauli matrices as $\gamma^{0}=-i\tau_{2}$\, ,  $\gamma^{1}=\tau_{1} $\, , and $ \psi^{\scriptscriptstyle T}=(\psi_{1} , \psi_{2}) \, ,\, \overline{\psi}= (\overline{\psi}_{1}, \overline{\psi}_{2})$ , this reads
\begin{equation}\label{eq: 03}
\mathcal{L}(t)=-\int_x\!\!\big{\lbrace}\overline{\psi}_{2}(\partial_{t}+\partial_{x}) \psi_{1}-\overline{\psi}_1(\partial_{t}-\partial_{x})\psi_{2}-2g\overline{\psi}_{2}\overline{\psi}_{1}\psi_{2}\psi_{1}\big{\rbrace}\; .
\end{equation}
We emphasize that the Minkowski action $S_{\scriptscriptstyle M}=-iS$ is not obtained by analytic continuation from the euclidean action. The Grassmann functional is not changed, the formulation with $S_{\scriptscriptstyle M}$ is just a different notation. In $S_{\scriptscriptstyle M}$ the interaction term is multiplied by $-ig /2$. This corresponds to an imaginary coupling in the Thirring model. Nevertheless we will see that the time evolution of this model is unitary for $g=1$. The Grassmann variables $\psi_{\gamma}$ and $\overline{\psi}_{\gamma}$ are independent. We deal with a real Grassmann algebra and therefore real wave functions. A complex structure will be introduced later.

For a well defined discrete formulation we choose for $t$ and $x$ a quadratic lattice with lattice distance $\varepsilon$. For a discretized version of our model we first consider the action
 \begin{equation}\label{eq: 07}
 S=\sum_{t\,\textup{even}}  \bigl[\mathcal{L}_\textup{kin}(t)+\mathcal{L}_\textup{int}(t)\bigr] \; ,
 \end{equation}
with
\begin{equation}\label{eq: 08}
\begin{split}
\mathcal{L}_\textup{kin}(t)\!=\!-\!\!&\sum_{x\,\textup{even}}\big[\varphi(t+\varepsilon, x+\varepsilon)\varphi(t,x) \\
&+\varphi(t+\varepsilon,x)\varphi(t, x+\varepsilon) \\ 
&+\varphi(t+2\varepsilon, x)\varphi(t+\varepsilon, x-\varepsilon)\\
&+\varphi(t+2\varepsilon, x-\varepsilon)\varphi(t+\varepsilon, x)\big]\; ,
\end{split}
\end{equation}
and
\begin{equation}\label{eq: 09}
\begin{split}
\mathcal{L}_\textup{int}(t)=g\!\!\sum_{x\,\textup{even}}[\varphi(t+\varepsilon, x+\varepsilon)\varphi(t+\varepsilon,x)\varphi(t, x+\varepsilon)\varphi(t,x)  \\ +\varphi(t+2\varepsilon, x-\varepsilon)\varphi(t+2\varepsilon, x)\varphi(t+\varepsilon, x-\varepsilon)\varphi(t+\varepsilon, x)]\; .
\end{split}
\end{equation}
The discrete action is an element of a real Grassmann algebra with a single Grassmann variable $\varphi(t,x)$ on each point of the lattice. The sum runs only over a coarse grained lattice with even $t$ and even $x$, or $m_{t}$, $m_{x}$ even integers.
We take periodic boundary conditions in $x$, with an even total number of $x$ points.

\bigskip
\noindent
{\bf Continuum limit}
\medskip

We introduce lattice derivatives by
\begin{align}\label{eq: 10}
(\partial_{t}+\partial_{x}) \varphi(t+\varepsilon, x+\varepsilon)=\dfrac{1}{2\varepsilon}\big[\varphi(t+2\varepsilon, x+2\varepsilon)-\varphi(t,x)\big]\;, \nn \\
 (\partial_{t}-\partial_{x}) \varphi(t+\varepsilon, x)=\dfrac{1}{2\varepsilon}\big[\varphi(t+2\varepsilon, x-\varepsilon)-\varphi(t,x+\varepsilon)\big]\;.
\end{align}
The kinetic term, 
\begin{equation}\label{eq: 11} 
\begin{split}
\mathcal{L}_\textup{kin}(t)=&2\varepsilon\!\!\sum_{x\, \textup{even}}\!\!\!\big{\lbrace} \varphi(t+\varepsilon, x+\varepsilon)(\partial_{t}+\partial_{x})\varphi(t+\varepsilon, x+\varepsilon) \\ &+\varphi(t+\varepsilon, x)(\partial_{t}-\partial_{x})\varphi(t+\varepsilon, x)\big{\rbrace}\; ,
\end{split}
\end{equation}
describes right movers on the sublattice with $m_{t}+m_{x}$ even, and left movers on the sublattice with $m_{t}+m_{x}$ odd.
For $t$, $x$ both even we choose the naming conventions and normalization
\begin{align}\label{eq: 12}
\varphi(t, x)=\sqrt{2\varepsilon}\,\psi_{1}(t,x)\; , \notag \\ \varphi(t, x+\varepsilon)=\sqrt{2\varepsilon}\,\psi_{2}(t,x+\varepsilon) \; , \notag \\
\varphi(t+\varepsilon, x)=\sqrt{2\varepsilon}\;\overline{\psi}_{1}(t+\varepsilon,x)\;, \notag \\ \varphi(t+\varepsilon , x+\varepsilon)=-\sqrt{2\varepsilon}\;\overline{\psi}_{2}(t+\varepsilon,x+\varepsilon)\; ,
\end{align}
such that
\begin{equation}\label{eq: 13}
\begin{split}
\!\mathcal{L}_\textup{kin}(t)\!=&\!-4\varepsilon^{2}\!\sum_{x\, \textup{even}}\\
&\bigl{\lbrace}\overline{\psi}_{2}(t+\varepsilon,x   +\varepsilon)(\partial_{t}+\partial_{x})\psi_{1}(t+\varepsilon,x+\varepsilon)  \\
&-\overline{\psi}_{1}(t+\varepsilon,x)(\partial_{t}-\partial_{x})\psi_{2}(t+\varepsilon, x)\bigr{\rbrace} \; .
\end{split}
\end{equation}

In the continuum limit the lattice derivatives become partial derivatives and $\sum_{t,\textup{even}}=(2\varepsilon)^{-1} \int\!dt$ , $\sum_{x,\textup{even}}\!\!=(2\varepsilon)^{-1} \int\! dx$ .
One recovers the derivative term in the action~\eqref{eq: 03}. In terms of $\psi$, $\overline{\psi}$ the interaction term~\eqref{eq: 09} reads
\begin{equation}\label{eq: 14}
\begin{split}
\mathcal{L}_\textup{int}(t)=&-4g\varepsilon^{2}\sum_{x,\textup{even}}\big{\lbrace}\overline{\psi}_{2}(t+\varepsilon,x+\varepsilon)\\ 
&\times\overline{\psi}_{1}(t +\varepsilon,x)\psi_{2}(t, x+\varepsilon)\psi_{1}(t,x) \phantom{\big{\rbrace}}\\  
&+\overline{\psi}_{2}(t+\varepsilon,x-\varepsilon)\overline{\psi}_{1}(t 
+\varepsilon,x)\phantom{\big{\rbrace}}\\ 
&\times\psi_{2}(t+2\varepsilon, x-\varepsilon)\psi_{1}(t+2\varepsilon, x)\big{\rbrace} \;.\!\!\!\!
\end{split}
\end{equation}
We find  the interaction term of eq.~\eqref{eq: 03} in the continuum limit.

\noindent
{\bf Local factors}
\medskip

For the action~\eqref{eq: 07}-\eqref{eq: 09} the weight function $\exp\big{\lbrace}-S[\varphi]\big{\rbrace}$ can be written as a product of local factors $\tilde{\mathcal{K}}(t)$ using
\begin{equation}\label{eq: 15}
\exp\lbrace -S\rbrace=\prod_{\substack{t, \textup{even}}} \exp \big{\lbrace}-{\mathcal{L}}(t)\big{\rbrace}=\prod_{t} \tilde{ \mathcal{K}}(t) \; ,
\end{equation}
where we take  for $ t $ even
\begin{equation}\label{eq: 16}
\exp\big{\lbrace}-\mathcal{L}(t)\big{\rbrace}=\tilde{\mathcal{K}}(t+\varepsilon)\tilde{\mathcal{K}}(t) \; .
\end{equation}
These local factors take the form
\begin{equation}\label{eq: 17}
\begin{split}
\tilde{\mathcal{K}}(t)=&\prod_{\substack{x, \textup{even}}} \big{\lbrace}1+\varphi(t+\varepsilon, x+\varepsilon)\varphi(t,x) \\ 
&+\varphi(t+\varepsilon, x)\varphi(t, x+\varepsilon)  \\
\;+(1-&g)\varphi(t+\varepsilon, x+\varepsilon)\varphi(t+\varepsilon, x)\varphi(t, x+\varepsilon)\varphi(t, x)\!\big{\rbrace}
\end{split}
\end{equation}
and
\begin{equation}\label{eq: 18} 
\begin{split}
\tilde{\mathcal{K}}(t+\varepsilon)=&\prod_{\substack{x, \textup{even}}} \big{\lbrace}1+\varphi(t+2\varepsilon, x-\varepsilon)\varphi(t+\varepsilon,x) \\
&+\varphi(t +2\varepsilon, x)\varphi(t+\varepsilon, x-\varepsilon) \\
&+(1-g)\varphi(t+2\varepsilon, x-\varepsilon)\varphi(t+2\varepsilon, x)\\
&\times\varphi(t+\varepsilon, x-\varepsilon)\varphi(t+\varepsilon, x)\big{\rbrace}\;  .
\end{split}
\end{equation}

Each factor for a given $x$ in $\tilde{\mathcal{K}}(t)$ for $t$ even involves only four Grassmann variables in a block of four lattice sites. The quadratic term multiplies variables at opposite ends of the diagonals in the block, and the interaction term involves all four variables in the block. For odd $t$ the structure is the same, only the blocks are shifted by one place in $x$ and $t$. As a consequence, every block in eq.~\eqref{eq: 17} has precisely one common variable with the two diagonal neighboring blocks from eq.~\eqref{eq: 18}.

The block structure makes no difference between space and time. We may view the model ~\eqref{eq: 07}-\eqref{eq: 09} as a euclidean fermion model on a two-dimensional square lattice. The choice of the time-direction is arbitrary at this stage. Time is singled out only by the direction in which we study the evolution. The Lorentz symmetry of the action~\eqref{eq: 02},~\eqref{eq: 03} with the characteristic difference in signature for space and time is not yet directly visible in the model~\eqref{eq: 07}-\eqref{eq: 09}, even though the diagonal structure may be seen already as a hint for light cones. As we have seen, it obtains in the continuum limit.

\bigskip
\noindent
\section{Cellular automata for fermions}\label{sec: 04}
\medskip

For given local factors of a discrete fermionic theory we can extract the step evolution operator by employing the relations~\eqref{eq: A17}~\eqref{eq: A20}. We first show that for suitable choices of $g$ the model ~\eqref{eq: 07}-\eqref{eq: 09} describes free fermions. We then move on to a somewhat more complex fermion model which describes the cellular automaton with the rules discussed in sect.~\ref{sec: 02} for the red and green particles.

\bigskip
\noindent
{\bf Free fermions}
\medskip

From eqs. ~\eqref{eq: 15}-~\eqref{eq: 18} we can extract the step evolution operators for the different blocks defined by appropriate pairs of positions. For each block it is a $4  \times 4$ matrix. For even $t$ the expansion ~\eqref{eq: A17} yields with eq.~\eqref{eq: A12}
\begin{equation}\label{eq: 19}
\widehat{S}(t,x)=\begin{pmatrix}
g\!-\!1\negthickspace&\phantom{-}0&\phantom{-}0&\phantom{-}0\\
0&\phantom{-}0&-1&\phantom{-}0\\
0&-1&\phantom{-}0&\phantom{-}0\\
0&\phantom{-}0&\phantom{-}0&\phantom{-}1
\end{pmatrix} \; ,
\end{equation}
while the expansion ~\eqref{eq: A20} implies
\begin{equation}\label{eq: 20}
\widehat{S}(t+\varepsilon,x)=\begin{pmatrix}
1&\phantom{-}0&\phantom{-}0&0\\
0&\phantom{-}0&\phantom{-}1&0\\
0&\phantom{-}1&\phantom{-}0&0\\
0&\phantom{-}0&\phantom{-}0&\;g\!-\!1
\end{pmatrix} \; .
\end{equation}
For $g=2$ one finds that $\widehat{S}\,(t+\varepsilon,x)$ in eq.~\eqref{eq: 20} is indeed the switch operator ~\eqref{eq: A08} employed for the first cellular automaton discussed in sect.~\ref{sec: 02}. It involves pairs at $x$ and $x-\varepsilon$ (for even $x$, even $t$, and therefore odd $t+\varepsilon$), as appropriate. For $t$ even the operator $\widehat{S}\,(t,x)$ in eq.~\eqref{eq: 19} acts on pairs at $x$ and $x+\varepsilon$, again in accordance with the cellular automaton.
The only difference concerns the minus signs in the off-diagonal elements. These minus signs have no physical significance. They can be absorbed by a different definition of Grassmann variables. Indeed, a variable transformation $\varphi(t,x)\rightarrow -\varphi(t,x)$ \,for\, $t=m_{t}\,\varepsilon \, , \,\, m_{t}={1\!,2} \bmod 4$ , changes the sign of the off-diagonal elements in eq.~\eqref{eq: 19}.
 
For $g=0$ one recovers freely propagating fermions. The operators ~\eqref{eq: 19}~\eqref{eq: 20} have eigenvalues $\pm 1$ . The product of these operators over all $x$ produces a simple step evolution operator. All particles on the sub-lattice with even $m_{t}+m_{x}$ move one place to the right if $m_{t}$ increases by one unit, while all particles on the other sublattice with odd $m_{t}+m_{x}$ move one place to the left. 
Thus $\psi_{1}, \overline{\psi}_{1}$ describe "right movers", and $\psi_{2}, \overline{\psi}_{2}$ "left-movers". 
 
For identical particles the model with $g=2$ is actually the same as the free model for $g=0$. These two values can be mapped onto each other by a change of sign of Grassmann basis functions. This can also be seen from Fig.~\ref{Fig.1}. Instead of moving the states $(1,1)$ and $(0,0)$ in the blocks by one position in $t$, corresponding to vertical black or red lines, we can also exchange the positions at $t+\varepsilon$, producing instead black or red crosses. As a result, all filled or empty positions move on diagonals, being right -or left - movers according to the appropriate sublattice. For identical particles Fig.~\ref{Fig.1} does not describe a non-trivial scattering.
We could equally well draw lines for unperturbed left and right movers. This can also be seen in the continuum limit. We recall that $\overline{\psi}(t+\varepsilon,x)$ is related to $\overline{\varphi}(t+\varepsilon, x)$ by eq.~\eqref{eq: 12}, while $\psi(t,x)$ corresponds to $\varphi(t,x)$. For sufficiently smooth wave functions $\varphi(t,x)$, for which a continuum limit applies, one would like to identify $\overline{\psi}_{\gamma}(t+\varepsilon,x)$ with variables $ \pm\,\psi_{\gamma}(t,x)$. In this limit the interaction term vanishes since only two distinct Grassmann variables are available at every point $(t,x)$.

\bigskip
\noindent
{\bf Dirac spinors}
\medskip

A fermion model with non trivial interaction in the continuum limit can be obtained if the fermions carry different colors.
If for Fig.~\ref{Fig.1} one particle is red, the other is green, they can be distinguished and the figure describes scattering due to interaction. We therefore consider next on each site of the lattice two different Grassmann variables. We may denote the two "colors" by $\varphi_{\scriptscriptstyle R}(t, x)$ for the red particles and $\varphi_{\scriptscriptstyle I}(t, x)$ for the green particles. The kinetic term ~\eqref{eq: 08} becomes a sum of two kinetic terms, one for $\varphi_{\scriptscriptstyle R}$ and the other for $\varphi_{\scriptscriptstyle I}$.

We can introduce a complex structure by defining the complex Grassmann variable
\begin{equation}\label{eq: 21}
\varphi(t,x)=\varphi_{\scriptscriptstyle R}(t, x)+i\varphi_{\scriptscriptstyle I}(t, x) \; .
\end{equation}
Correspondingly, on the level of $\psi$, $\overline{\psi}$ in eq.~\eqref{eq: 12} we define complex Grassmann variables by
\begin{align}\label{eq: 22}
\psi_{\eta}(t,x)=\psi_{\eta, {\scriptscriptstyle R}}(t,x)+i\psi_{\eta, {\scriptscriptstyle I}}(t,x) \; , \nn \\
\overline{\psi}_{\eta}(t,x)=\overline{\psi}_{\eta, {\scriptscriptstyle R}}(t,x)-i\overline{\psi}_{\eta, {\scriptscriptstyle I}}(t,x) \; .
\end{align}
The kinetic term ~\eqref{eq: 13} retains its form, except that $\psi_{\eta}$ and $\overline{\psi}_{\eta}$ are now complex Grassmann variables and the real part of $\mathcal{L}_\textup{kin}(t)$ is taken. The continuum limit yields the kinetic term in eq.~\eqref{eq: 02}, now with complex Grassmann variables $\psi$, $\overline{\psi}$ as appropriate for Dirac spinors. An interaction term becomes possible in the continuum limit. If it is given by eq.~\eqref{eq: 02} one realizes a type of Thirring model. For real $g$ it has an "imaginary coupling".

\bigskip
\noindent
{\bf Discrete model for interacting fermions}
\medskip

For a discrete formulation of Thirring type models we need to add an interaction term whose continuum limit yields the interaction term in eq.~\eqref{eq: 02}, with $\psi$, $\overline{\psi}$ now complex Grassmann variables. We propose local factors $\tilde{\mathcal{K}}(t,x)$ for the blocks of four lattice sites at even $t$, $x$, with $\alpha=R,I$\;, 
\begin{equation}\label{eq: 23}
\tilde{\mathcal{K}}(t, x)=\exp\big{\lbrace}-\tilde{\mathcal{L}}_{\textup{kin}}(t,x)\big{\rbrace}+\tilde{\mathcal{K}}_{\textup{int}}(t,x) \; ,
\end{equation}
with
\begin{equation}\label{eq: 23A}
\begin{split}
\tilde{\mathcal{L}}_{\textup{kin}}(t,x)=-\bigg{\lbrace}\sum_{\alpha}\bigl[\varphi_{\alpha}(t+\varepsilon,x+\varepsilon)\varphi_{\alpha}(t,x)\\
+\varphi_{\alpha}(t+\varepsilon,x)\varphi_{\alpha}(t, x+\varepsilon)\bigr]\bigg{\rbrace} \;.
\end{split}
\end{equation}

The interaction part reads
\begin{equation}\label{eq: 24}
\begin{split}
\tilde{\mathcal{K}}_{\textup{int}}(t,x)&=(\zeta'_{1}\zeta'_{4}-\zeta'_{2}\zeta'_{3})(\zeta_{1}\zeta_{4}-\zeta_{2}\zeta_{3})  \\
&-(\zeta'_{1}\zeta'_{3}+\zeta'_{2}\zeta'_{4})(\zeta_{1}\zeta_{3}+\zeta_{2}\zeta_{4}) \; ,
\end{split}
\end{equation}
where the Grassmann variables on the four sites of the block are given by

\begin{align}\label{eq: 25}
\zeta_{1}\!=\varphi_{\scriptscriptstyle R}(t,x)\!=\sqrt{2\varepsilon}\,\psi_{1{\scriptscriptstyle R}}(t,x)\, , \nn \\
\zeta_{2}\!=\varphi_{\scriptscriptstyle I}(t,x)\!=\sqrt{2\varepsilon}\,\psi_{1{\scriptscriptstyle I}}(t,x)\, ,\nn \\
\zeta_{3}=\varphi_{\scriptscriptstyle R}(t,x+\varepsilon)=\sqrt{2\varepsilon}\,\psi_{2{\scriptscriptstyle R}}(t,x+\varepsilon)\, , \nn \\ 
\zeta_{4}=\varphi_{\scriptscriptstyle I}(t,x+\varepsilon)=\sqrt{2\varepsilon}\,\psi_{2{\scriptscriptstyle I}}(t,x+\varepsilon)\; ,
\end{align}
and
\begin{align}\label{eq:25a}
\zeta'_{1}=\varphi_{\scriptscriptstyle R}(t+\varepsilon,x)=\sqrt{2\varepsilon}\,\overline{\psi}_{1{\scriptscriptstyle R}}(t+\varepsilon,x)\, ,\nn \\ 
\zeta'_{2}=\varphi_{\scriptscriptstyle I}(t+\varepsilon,x)=\sqrt{2\varepsilon}\,\overline{\psi}_{1{\scriptscriptstyle I}}(t+\varepsilon,x)\, ,\nn \\
\zeta'_{3}=\varphi_{\scriptscriptstyle R}(t+\varepsilon,x+\varepsilon)=-\sqrt{2\varepsilon}\,\overline{\psi}_{2{\scriptscriptstyle R}}(t+\varepsilon,x+\varepsilon)\, , \nn \\
\zeta'_{4}=\varphi_{\scriptscriptstyle I}(t+\varepsilon,x+\varepsilon)=-\sqrt{2\varepsilon}\,\overline{\psi}_{2{\scriptscriptstyle I}}(t+\varepsilon,x+\varepsilon)\;. 
\end{align}
Omitting the differences between the different positions one finds in terms of the complex doublets $\psi$ and $\overline{\psi}$
\begin{equation}\label{eq: 26}
\tilde{\mathcal{K}}_{\textup{int}}(t,x)=\varepsilon^{2}(\overline{\psi}\gamma^{\mu}\psi)^{*}(\overline{\psi}\gamma_{\mu}\psi)\;.
\end{equation}
For odd $t+\varepsilon$ we use for $\tilde{\mathcal{K}}(t+\varepsilon)$ the same from~\eqref{eq: 23}, with $t$ replaced by $t+\varepsilon$ and $x$ replaced by $x+\varepsilon$. Equivalently, we could also make the shift $x\rightarrow x-\varepsilon$ \;.

We next establish that the local factors~\eqref{eq: 23} correspond to the discretization of a particular Thirring model. Writing the block local factor in exponential form
\begin{equation}\label{eq: 27}
\tilde{\mathcal{K}}(t,x)=\exp\big{\lbrace}-\tilde{\mathcal{L}}(t,x)\big{\rbrace} \;,
\end{equation}
one has
\begin{equation}\label{eq: 28}
\tilde{\mathcal{L}}(t,x)=\tilde{\mathcal{L}}_{\textup{kin}}(t,x)-\tilde{\mathcal{K}}_{\textup{int}}-\Delta \;,
\end{equation}
where the "correction term" $\Delta$ obeys
\begin{equation}\label{eq: 30}
\Delta=\tilde{\mathcal{K}}_{\textup{int}}(\tilde{\mathcal{L}}_{\textup{kin}}+\frac{1}{2}\tilde{\mathcal{L}}^{2}_{\textup{kin}})-\frac{1}{2}\tilde{\mathcal{K}}^2_{\textup{int}}\;.
\end{equation}
Here we use that $\tilde{\mathcal{K}}_{\textup{int}}$ involves four Grassmann variables, and $\Delta$ at least six Grassmann variables, such that $\tilde{\mathcal{K}}^{3}_{\textup{int}}=0 \, , \, \tilde{\mathcal{K}}_{\textup{int}}\Delta=0\,,\, \Delta^{2}=0$ \;.

In the continuum limit the first two terms in eq.~\eqref{eq: 28} yield
\begin{equation}\label{eq: 31}
S=\int_{t,x}\!\big{\lbrace}-Re(\overline{\psi}\gamma^{\mu}\partial_{\mu}\psi)-\frac{1}{2}(\overline{\psi}\gamma^{\mu}\psi)^{*}(\overline{\psi}\gamma_{\mu}\psi)\big{\rbrace} \;.
\end{equation}
The term $\Delta$ vanishes in the continuum limit $\varepsilon\rightarrow0$\;, with
\begin{equation}\label{eq:31a}
\Delta\sim\varepsilon^{3}(\overline{\psi}\psi)^{3}\, , \quad \sum_{t,x}\Delta\sim\varepsilon\!\int_{t,x}\!(\overline{\psi}\psi)^{3} \;.
\end{equation}
Also the differences between the different positions vanish in the transition from eq.~\eqref{eq: 25} to eq.~\eqref{eq: 26}. Only the part~\eqref{eq: 31} remains. In the continuum limit one also wants to express $ \overline{\psi}(t,x)$ in terms of $\psi(t,x)$ since it originates from $\psi(t+\varepsilon,x)$. The identification reads
\begin{equation}\label{eq: 32}
\overline{\psi}(t,x)=\psi^{\dagger}(t,x)\gamma^{0} \;.
\end{equation}
With this identification $\int\! \overline{\psi}\gamma^{\mu}\partial_{\mu}\psi$ is real by use of partial integration, and $(\overline{\psi}\gamma^{\mu}\psi)^{*}=-\overline{\psi}\gamma^{\mu}\psi$\;.
The continuum limit yields indeed the Thirring model~\eqref{eq: 02} with $g=1$, now with complex Grassmann variables $\psi_{1}, \psi_{2}$\,.

\bigskip
\noindent
{\bf Cellular automaton for interacting fermions}
\medskip

For the particular Thirring type model with $g=1$ we can extract the step evolution operator from the local factor~\eqref{eq: 23}. For each block it is a $16\times 16$ matrix. The kinetic part alone (first term in eq.~\eqref{eq: 23}) moves all four configurations of particle numbers from the lower left corner $(t,x)$ of the block to the upper right corner $(t+\varepsilon, x+\varepsilon)$\,. 
Similarly, it transports all four configurations in the lower right corner $(t, x+\varepsilon)$ to the upper left corner $(t+\varepsilon, x)$. This would be a generalized switch operator. In the presence of the interaction term, however, this behavior is modified if both lower corners of the block are occupied by a single particle. This concerns the products of variables $\zeta_{1}\zeta_{3}, \, \zeta_{1}\zeta_{4}, \, \zeta_{2}\zeta_{3}$ and $\zeta_{2}\zeta_{4}$. In this sector the kinetic term contributes a factor 
\begin{equation}\label{eq: 33}
\begin{split}
\tilde{\mathcal{K}}_{\textup{kin}}&=(\zeta'_{1}\zeta_{3}+\zeta'_{2}\zeta_{4}+\zeta'_{3}\zeta_{1}+\zeta'_{4}\zeta_{2})^{2} \\ &=\zeta'_{1}\zeta'_{3}\zeta_{1}\zeta_{3}+\zeta'_{1}\zeta'_{4}\zeta_{2}\zeta_{3}+\zeta'_{2}\zeta'_{3}\zeta_{1}\zeta_{4}+\zeta'_{2}\zeta'_{4}\zeta_{2}\zeta_{4} \; ,
\end{split}
\end{equation}
as obtained from the expansion of the exponential and omitting contributions $\sim\zeta_{1}\zeta_{2}$ or $\sim\zeta_{3}\zeta_{4}$ \;. The interaction part contributes
\begin{equation}\label{eq: 34}
\begin{split}
\tilde{\mathcal{K}}_{\textup{int}}&=\zeta'_{1}\zeta'_{4}\zeta_{1}\zeta_{4}+\zeta'_{2}\zeta'_{3}\zeta_{2}\zeta_{3} 
-\zeta'_{1}\zeta'_{3}\zeta_{2}\zeta_{4}-\zeta'_{2}\zeta'_{4}\zeta_{1}\zeta_{3} \\
&\; \;\;-\zeta'_{1}\zeta'_{3}\zeta_{1}\zeta_{3}-\zeta'_{1}\zeta'_{4}\zeta_{2}\zeta_{3}
-\zeta'_{2}\zeta'_{3}\zeta_{1}\zeta_{4}-\zeta'_{2}\zeta'_{4}\zeta_{2}\zeta_{4} \;.
\end{split}
\end{equation}
The last two terms cancel $\tilde{\mathcal{K}}_{\textup{kin}}$, such that only the first two terms remain in this sector.

With a suitable choice of sign for the Grassmann basis functions the step evolution operator obtained from eqs.~\eqref{eq: A17}~\eqref{eq: A20} is a unique jump operator with one element equal to one in each line and column, and zeros otherwise. It describes a cellular automaton with the rules for the blocks
given by the ones for the cellular automaton with red and green particles specified in sect.~\ref{sec: 02}. The processes ~\ref{item: 1}.)~and~\ref{item: 2}) arise from $\tilde{\mathcal{K}}_{\textup{int}}+\tilde{\mathcal{K}}_{\textup{kin}}$ in eqs.\eqref{eq: 33},~\eqref{eq: 34}.

In summary, the discrete fermion model~\eqref{eq: 23}-\eqref{eq: 24} is equivalent to a cellular automaton. Its continuum limit is a particular type of Thirring model. It is invariant under Lorentz-transformations, and shows all features of a quantum field theory with interactions. The evolution of the wave function is unitary, as established by the explicit construction of a unitary evolution operator. Indeed, the orthogonal step evolution operator in the real formulation becomes a unitary evolution in the presence of a complex structure. The norm of the wave function is preserved.

In the discrete formulation the complex multi-particle wave functions form a finite-dimensional Hilbert space. Linear combinations and the scalar product obtain directly from corresponding quantities for the real wave functions $q_{\tau}(t)$. For a well defined continuum limit the number of different occupation numbers at fixed $t$ goes to infinity. The Hilbert space becomes infinite-dimensional. All properties of the quantum many body systems corresponding to the quantum field theory are faithfully represented by the evolution of the wave function for the probabilistic cellular automaton.

\bigskip
\noindent
\section{Symmetries, ground states and \\ dynamics}\label{sec: 05}

Many properties of the dynamics generated by the cellular automaton can be understood in terms of the dynamics of the particular Thirring model. This concerns symmetries, possible ground states and spontaneous symmetry breaking, and the existence of soliton solutions.

\medskip
\noindent
{\bf Symmetries}
\medskip

Many symmetries of the continuous Thirring-type  model~\eqref{eq: 02} are already present in the discrete fermion model~\eqref{eq: 23}-\eqref{eq: 24} and can be seen directly for the cellular automaton. Chiral transformations rotate the two colors into each other, separately for right movers and left movers. Infinitesimal chiral transformations act on the complex fields $\psi_{1}$ and $\psi_{2}$ as separate phase transformations. On the Grassmann variables~\eqref{eq: 25} infinitesimal chiral transformations act as
\begin{align}\label{51}
\delta\zeta_{1}=\xi_{\scriptscriptstyle +}\zeta_{2} \quad,\quad \delta\zeta_{2}=-\xi_{\scriptscriptstyle +}\zeta_{1}\;, \nn \\
\delta\zeta_{3}=\xi_{\scriptscriptstyle -}\zeta_{4} \quad,\quad \delta\zeta_{4}=-\xi_{\scriptscriptstyle -}\zeta_{3}\;, \nn \\
\delta\zeta'_{1}=\xi_{\scriptscriptstyle -}\zeta'_{2} \quad,\quad \delta\zeta'_{2}=-\xi_{\scriptscriptstyle -}\zeta_{1}\;, \nn \\
\delta\zeta'_{3}=\xi_{\scriptscriptstyle +}\zeta'_{4} \quad,\quad \delta\zeta'_{4}=-\xi_{\scriptscriptstyle +}\zeta_{3}\;.
\end{align}
Both $\tilde{\mathcal{L}}_{\textup{kin}}$ and $\tilde{\mathcal{K}}_{\textup{int}}$ in eqs.~\eqref{eq: 23A},~\eqref{eq: 24} are invariant. The chiral symmetries imply the separate conservation of the numbers of right movers and left movers, which is clearly a property of the cellular automaton. 

Discrete symmetries comprise parity (reflection of $x$) and time reversal (reflection of $t$). Charge conjugation corresponds to complex conjugation of $\psi$, or a change of sign of the Grassmann variables $\varphi_{\eta, {\scriptscriptstyle I}}$. For the cellular automaton this means that the number of green particles can only change in units of two. The exchange of red and green particles in the cellular automaton corresponds to invariance under the discrete symmetry $\psi\rightarrow i\psi$, $\overline{\psi}\rightarrow -i\overline{\psi}$\;.

\bigskip
\noindent
{\bf Ground states}
\medskip

We define a ground state as a time-invariant state. For a ground state configuration at $t$ the configuration at $t+2\varepsilon $ is the same as for $t$. Because of the modulo two properties of our model for even and odd $t$ we take a coarse grained approach for the definition of time invariance. It guarantees that the state for all even $t+2n\varepsilon$ is the same as at~$t$.

There a several possible ground states according to this definition. First of all, we have the totally empty state where all occupation numbers vanish or all Ising spins take negative values. A single one-particle excitation in this ground state takes at $t_{\textup{in}}$ a single occupation number different from zero. One has $n_{\alpha}(t_{\textup{in}},x)=1$ for one particular $x$ and one particular $\alpha$, and keeps all other occupation numbers at zero. The single particle will propagate for increasing $t$ as a right mover or a left mover without scattering, depending on $x$  being even or odd. The ground state is stable in the sense that for later $t$ it remains the same as at $t_{\textup{in}}$ for all locations $x$ except for the position of the particle. A general one-particle wave function $q_{\alpha}(t,x)$ associates to each one-particle configuration the probability $q_{\alpha}^{2}(t,x)$~. 

A second ground state is the totally filled state, with all $n_{\alpha}(t,x)=1$\;, or all Ising spins positive. The single particle excitation is now a hole, corresponding to $n_{\alpha}(t_{\textup{in}},x)=0$ for a single $x$ and $\alpha$.
The hole propagates in the same way as a particle in the first ground state. Particle-hole symmetry maps the two ground states and the two one-particle states onto each other. These two ground states preserve the chiral symmetry of the action.

\begin{figure}[h]
\resizebox{12pc}{!}{%
\begin{tikzpicture}
\draw [blue] (0,0) rectangle (13,13);
\draw[black!60!green, line width=3pt](0,1) --(0.5,0.5) --++(1,1)--++(1,-1) --++(1,1)--++(1,-1) --++(1,1)--++(1,-1) --++(1,1)--++(1,-1) --++(1,1)--++(1,-1) --++(1,1)--++(1,-1) --++(0.5,0.5);
\draw[black!60!green, line width=3pt](0,3) --(0.5,2.5) --++(1,1)--++(1,-1) --++(1,1)--++(1,-1) --++(1,1)--++(1,-1) --++(1,1)--++(1,-1) --++(1,1)--++(1,-1) --++(1,1)--++(1,-1) --++(0.5,0.5);
\draw[black!60!green, line width=3pt](0,5) --(0.5,4.5) --++(1,1)--++(1,-1) --++(1,1)--++(1,-1) --++(1,1)--++(1,-1) --++(1,1)--++(1,-1) --++(1,1)--++(1,-1) --++(1,1)--++(1,-1) --++(0.5,0.5);
\draw[black!60!green, line width=3pt](0,7) --(0.5,6.5) --++(1,1)--++(1,-1) --++(1,1)--++(1,-1) --++(1,1)--++(1,-1) --++(1,1)--++(1,-1) --++(1,1)--++(1,-1) --++(1,1)--++(1,-1) --++(0.5,0.5);
\draw[black!60!green, line width=3pt](0,9) --(0.5,8.5) --++(1,1)--++(1,-1) --++(1,1)--++(1,-1) --++(1,1)--++(1,-1) --++(1,1)--++(1,-1) --++(1,1)--++(1,-1) --++(1,1)--++(1,-1) --++(0.5,0.5);
\draw[black!60!green, line width=3pt](0,11) --(0.5,10.5) --++(1,1)--++(1,-1) --++(1,1)--++(1,-1) --++(1,1)--++(1,-1) --++(1,1)--++(1,-1) --++(1,1)--++(1,-1) --++(1,1)--++(1,-1) --++(0.5,0.5);
\draw[black!60!green, line width=3pt] (0,13)--(0.5,12.5) --(1,13) ++(1,0)--++(0.5,-0.5)--++(0.5,0.5) ++(1,0)--++(0.5,-0.5)--++(0.5,0.5) ++(1,0)--++(0.5,-0.5)--++(0.5,0.5) ++(1,0)--++(0.5,-0.5)--++(0.5,0.5) ++(1,0)--++(0.5,-0.5)--++(0.5,0.5) ++(1,0)--++(0.5,-0.5)--++(0.5,0.5);
\draw[red!85](0,0) --(0.5,0.5) -- (1,0) ++(1,0)--++(0.5,0.5) --++(0.5,-0.5) ++(1,0)--++(0.5,0.5) --++(0.5,-0.5) ++(1,0)--++(0.5,0.5) --++(0.5,-0.5) ++(1,0)--++(0.5,0.5) --++(0.5,-0.5) ++(1,0)--++(0.5,0.5) --++(0.5,-0.5) ++(1,0)--++(0.5,0.5) --++(0.5,-0.5) ;
\draw[red!85] (0,2) --(0.5,2.5) --++(1,-1)-- ++(1,1) --++(1,-1)-- ++(1,1) --++(1,-1)-- ++(1,1) --++(1,-1)-- ++(1,1) --++(1,-1)-- ++(1,1) --++(1,-1)-- ++(1,1) --++(0.5,-0.5);
\draw[red!85] (0,4) --(0.5,4.5) --++(1,-1)-- ++(1,1) --++(1,-1)-- ++(1,1) --++(1,-1)-- ++(1,1) --++(1,-1)-- ++(1,1) --++(1,-1)-- ++(1,1) --++(1,-1)-- ++(1,1) --++(0.5,-0.5);
\draw[red!85] (0,6) --(0.5,6.5) --++(1,-1)-- ++(1,1) --++(1,-1)-- ++(1,1) --++(1,-1)-- ++(1,1) --++(1,-1)-- ++(1,1) --++(1,-1)-- ++(1,1) --++(1,-1)-- ++(1,1) --++(0.5,-0.5);
\draw[red!85] (0,8) --(0.5,8.5) --++(1,-1)-- ++(1,1) --++(1,-1)-- ++(1,1) --++(1,-1)-- ++(1,1) --++(1,-1)-- ++(1,1) --++(1,-1)-- ++(1,1) --++(1,-1)-- ++(1,1) --++(0.5,-0.5);
\draw[red!85] (0,10) --(0.5,10.5) --++(1,-1)-- ++(1,1) --++(1,-1)-- ++(1,1) --++(1,-1)-- ++(1,1) --++(1,-1)-- ++(1,1) --++(1,-1)-- ++(1,1) --++(1,-1)-- ++(1,1) --++(0.5,-0.5);
\draw[red!85] (0,12) --(0.5,12.5) --++(1,-1)-- ++(1,1) --++(1,-1)-- ++(1,1) --++(1,-1)-- ++(1,1) --++(1,-1)-- ++(1,1) --++(1,-1)-- ++(1,1) --++(1,-1)-- ++(1,1) --++(0.5,-0.5);
\end{tikzpicture}}
\caption{Half filled ground state (A). Green and red lines link sites occupied by a single particle of the corresponding color.}\label{Fig.3}
\end{figure}
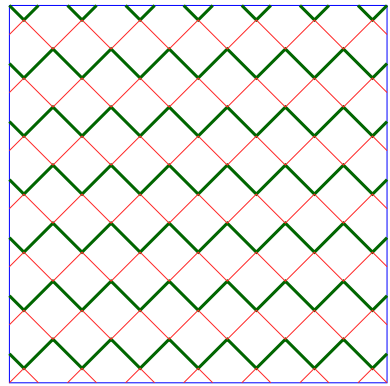

Another type of ground states are the half-filled states. They have precisely one particle at each location $x$
\begin{equation}\label{eq: GS1}
n_{\scriptscriptstyle R}(t,x)+n_{\scriptscriptstyle I}(t,x)=1 \;.
\end{equation}
The colors  do change, however, from $t$ to $t+\varepsilon$. They are again the same at $t+2\varepsilon$. For a first ground state of this type we take at $t_{\textup{in}}$ only red particles, $n_{\scriptscriptstyle R}(t_{\textup{in}},x)=1$\,, $n_{\scriptscriptstyle I}(t_{\textup{in}},x)=0$. The evolution is shown in Fig.~\ref{Fig.3}. At $t+\varepsilon$ all particles are green, and at $t+2\varepsilon$ red again. A second possible ground state of this type has only green particles at $t_{\textup{in}}$.
Chiral symmetry transformations change red into green particles. The two half-filled ground states correspond to spontaneous chiral symmetry breaking. The symmetry of the ground state is reduced as compared to the symmetry of the action. We call this structure of the half-filled ground state "type A".

\begin{figure}[h]
\resizebox{12pc}{!}{%
\begin{tikzpicture}
\def\alfa{90} 
\begin{scope}[rotate around={\alfa:(7,7)}]
\draw [blue] (0,0) rectangle (13,13);
\draw[black!60!green, line width=3pt](0,1) --(0.5,0.5) --++(1,1)--++(1,-1) --++(1,1)--++(1,-1) --++(1,1)--++(1,-1) --++(1,1)--++(1,-1) --++(1,1)--++(1,-1) --++(1,1)--++(1,-1) --++(0.5,0.5);
\draw[black!60!green, line width=3pt](0,3) --(0.5,2.5) --++(1,1)--++(1,-1) --++(1,1)--++(1,-1) --++(1,1)--++(1,-1) --++(1,1)--++(1,-1) --++(1,1)--++(1,-1) --++(1,1)--++(1,-1) --++(0.5,0.5);
\draw[black!60!green, line width=3pt](0,5) --(0.5,4.5) --++(1,1)--++(1,-1) --++(1,1)--++(1,-1) --++(1,1)--++(1,-1) --++(1,1)--++(1,-1) --++(1,1)--++(1,-1) --++(1,1)--++(1,-1) --++(0.5,0.5);
\draw[black!60!green, line width=3pt](0,7) --(0.5,6.5) --++(1,1)--++(1,-1) --++(1,1)--++(1,-1) --++(1,1)--++(1,-1) --++(1,1)--++(1,-1) --++(1,1)--++(1,-1) --++(1,1)--++(1,-1) --++(0.5,0.5);
\draw[black!60!green, line width=3pt](0,9) --(0.5,8.5) --++(1,1)--++(1,-1) --++(1,1)--++(1,-1) --++(1,1)--++(1,-1) --++(1,1)--++(1,-1) --++(1,1)--++(1,-1) --++(1,1)--++(1,-1) --++(0.5,0.5);
\draw[black!60!green, line width=3pt](0,11) --(0.5,10.5) --++(1,1)--++(1,-1) --++(1,1)--++(1,-1) --++(1,1)--++(1,-1) --++(1,1)--++(1,-1) --++(1,1)--++(1,-1) --++(1,1)--++(1,-1) --++(0.5,0.5);
\draw[black!60!green, line width=3pt] (0,13)--(0.5,12.5) --(1,13) ++(1,0)--++(0.5,-0.5)--++(0.5,0.5) ++(1,0)--++(0.5,-0.5)--++(0.5,0.5) ++(1,0)--++(0.5,-0.5)--++(0.5,0.5) ++(1,0)--++(0.5,-0.5)--++(0.5,0.5) ++(1,0)--++(0.5,-0.5)--++(0.5,0.5) ++(1,0)--++(0.5,-0.5)--++(0.5,0.5);
\draw[red!85](0,0) --(0.5,0.5) -- (1,0) ++(1,0)--++(0.5,0.5) --++(0.5,-0.5) ++(1,0)--++(0.5,0.5) --++(0.5,-0.5) ++(1,0)--++(0.5,0.5) --++(0.5,-0.5) ++(1,0)--++(0.5,0.5) --++(0.5,-0.5) ++(1,0)--++(0.5,0.5) --++(0.5,-0.5) ++(1,0)--++(0.5,0.5) --++(0.5,-0.5) ;
\draw[red!85] (0,2) --(0.5,2.5) --++(1,-1)-- ++(1,1) --++(1,-1)-- ++(1,1) --++(1,-1)-- ++(1,1) --++(1,-1)-- ++(1,1) --++(1,-1)-- ++(1,1) --++(1,-1)-- ++(1,1) --++(0.5,-0.5);
\draw[red!85] (0,4) --(0.5,4.5) --++(1,-1)-- ++(1,1) --++(1,-1)-- ++(1,1) --++(1,-1)-- ++(1,1) --++(1,-1)-- ++(1,1) --++(1,-1)-- ++(1,1) --++(1,-1)-- ++(1,1) --++(0.5,-0.5);
\draw[red!85] (0,6) --(0.5,6.5) --++(1,-1)-- ++(1,1) --++(1,-1)-- ++(1,1) --++(1,-1)-- ++(1,1) --++(1,-1)-- ++(1,1) --++(1,-1)-- ++(1,1) --++(1,-1)-- ++(1,1) --++(0.5,-0.5);
\draw[red!85] (0,8) --(0.5,8.5) --++(1,-1)-- ++(1,1) --++(1,-1)-- ++(1,1) --++(1,-1)-- ++(1,1) --++(1,-1)-- ++(1,1) --++(1,-1)-- ++(1,1) --++(1,-1)-- ++(1,1) --++(0.5,-0.5);
\draw[red!85] (0,10) --(0.5,10.5) --++(1,-1)-- ++(1,1) --++(1,-1)-- ++(1,1) --++(1,-1)-- ++(1,1) --++(1,-1)-- ++(1,1) --++(1,-1)-- ++(1,1) --++(1,-1)-- ++(1,1) --++(0.5,-0.5);
\draw[red!85] (0,12) --(0.5,12.5) --++(1,-1)-- ++(1,1) --++(1,-1)-- ++(1,1) --++(1,-1)-- ++(1,1) --++(1,-1)-- ++(1,1) --++(1,-1)-- ++(1,1) --++(1,-1)-- ++(1,1) --++(0.5,-0.5);
\end{scope}
\end{tikzpicture}}
\caption{Half filled ground state (B). It can be obtained from Fig. \ref{Fig.3} by a $\pi/2$ - rotation.}\label{Fig.4}
\end{figure}
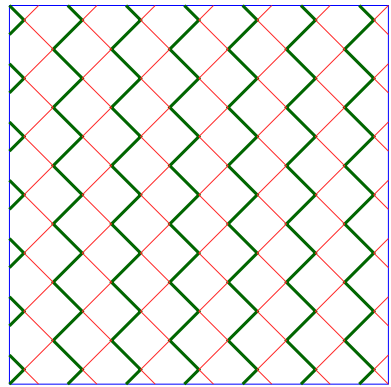

The condition~\eqref{eq: GS1} allows for many more different ground states. We may consider two ground states of "type B", for which at $t_{\textup{in}}$ red and green particles fill the positions in an alternating way. We despict this ground state in Fig.~\ref{Fig.4}. The other ground state of this type interchanges the red and green particles. As compared to the type A in Fig.~\ref{Fig.3}, the ground state of type B in Fig.~\ref{Fig.4} obtains by interchanging $x$ and $t$. 

\noindent

\bigskip
\noindent
{\bf Solitons}
\medskip

All the different ground states break some of the discrete symmetries of the action spontaneously. We may therefore have solutions where different regions in spacetime are characterized by different "vacua" or ground states. At the boundaries of these regions there will be "domain walls" or "kinks" or "solitons". There exist various solitons of this type, as determined by different vacua in the different regions. We only briefly discuss one type here.

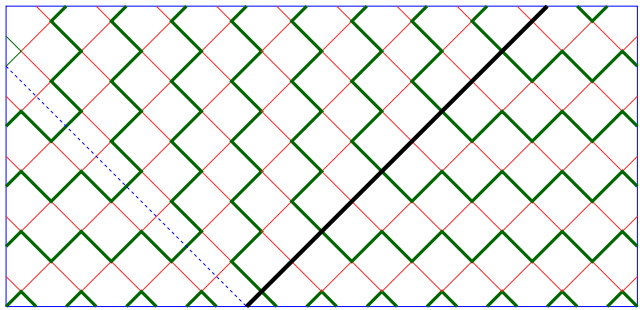
\begin{figure}[h]
\resizebox{20pc}{!}{%
\begin{tikzpicture}

\draw [blue] (0,0) rectangle (21,10);
\draw[red!85](0,1) --(0.5,0.5) --++(1,1)--++(1,-1) --++(1,1)--++(1,-1) --++(1,1)--++(1,-1) --++(1,1)--++(-1,1)--++(1,1)--++(-1,1)--++(1,1)--++(-1,1)--++(1,1)--++(-1,1)--++(1,1)--++(-0.5,0.5);
\draw[red!85](0,3) --(0.5,2.5) --++(1,1)--++(1,-1) --++(1,1)--++(1,-1) --++(1,1)--++(-1,1)--++(1,1)--++(-1,1)--++(1,1)--++(-1,1)--++(1,1)--++(-0.5,0.5);
\draw[red!85](0,5) --(0.5,4.5) --++(1,1)--++(1,-1) --++(1,1)--++(-1,1)--++(1,1)--++(-1,1)--++(1,1) --++(-0.5,0.5);
\draw[red!85](0,7) --(0.5,6.5) --++(1,1)--++(-1,1)--++(1,1)--++(-0.5,0.5);
\draw[black!60!green, line width=3pt](0,0) --(0.5,0.5) -- (1,0) ++(1,0)--++(0.5,0.5) --++(0.5,-0.5) ++(1,0)--++(0.5,0.5) --++(0.5,-0.5) ++(1,0)--++(0.5,0.5) --++(0.5,-0.5)  ; 
\draw[black!60!green, line width=3pt](9,0)--++(-1.5,+1.5)--++(1,1)--++(-1,1)--++(1,1)--++(-1,1)--++(1,1)--++(-1,1)--++(1,1)--++(-1,1)--++(0.5,0.5);
\draw[black!60!green, line width=3pt](10,0) --++(0.5,0.5)--++(0.5,-0.5) ++(1,0) --++(0.5,0.5)--++(0.5,-0.5) ++(1,0) --++(0.5,0.5) --++(0.5,-0.5) ++(1,0)--++(0.5,0.5) --++(0.5,-0.5) ++(1,0)--++(0.5,0.5) --++(0.5,-0.5) ++(1,0)--++(0.5,0.5)--++(0.5,-0.5) ;
\draw[black!60!green, line width=3pt] (0,2) --(0.5,2.5) --++(1,-1)-- ++(1,1) --++(1,-1)-- ++(1,1) --++(1,-1)-- ++(1,1)--++(-1,1)--++(1,1)--++(-1,1)--++(1,1)--++(-1,1)--++(1,1)--++(-1,1)--++(0.5,0.5);
\draw[black!60!green, line width=3pt] (0,4) --(0.5,4.5) --++(1,-1)-- ++(1,1) --++(1,-1)-- ++(1,1)--++(-1,1)--++(1,1)--++(-1,1)--++(1,1)--++(-1,1)--++(0.5,0.5);
\draw[black!60!green, line width=3pt] (0,6) --(0.5,6.5) --++(1,-1)-- ++(1,1)--++(-1,1)--++(1,1)--++(-1,1)--++(0.5,0.5);
\draw[black!60!green] (0,8) --(0.5,8.5)--++(-0.5,0.5);
\draw [red!85] (21,1)--++(-0.5,-0.5)--++(-1,1)--++(-1,-1)--++(-1,1)--++(-1,-1)--++(-1,1)--++(-1,-1)--++(-1,1)--++(-1,-1)--++(-1,1)--++(-1,-1)--++( -2,2)-- ++(1,1)--++(-1,1)-- ++(1,1)--++(-1,1)-- ++(1,1)--++(-1,1)--++(1,1)--++(-0.5,0.5);
\draw [red!85] (21,3)--++(-0.5,-0.5)--++(-1,1)--++(-1,-1)--++(-1,1)--++(-1,-1)--++(-1,1)--++(-1,-1)--++(-1,1)--++(-1,-1)--++( -2,2)-- ++(1,1)--++(-1,1)-- ++(1,1)--++(-1,1)--++(1,1)--++(-0.5,0.5);
\draw [red!85] (21,5)--++(-0.5,-0.5)--++(-1,1)--++(-1,-1)--++(-1,1)--++(-1,-1)--++(-1,1)--++(-1,-1)--++( -2,2)-- ++(1,1)--++(-1,1)-- ++(1,1)--++(-0.5,0.5);
\draw [red!85] (21,7)--++(-0.5,-0.5)--++(-1,1)--++(-1,-1)--++(-1,1)--++(-1,-1)--++( -2,2)-- ++(1,1)--++(-0.5,0.5);
\draw [red!85] (21,9)--++(-0.5,-0.5)--++(-1,1)--++(-1,-1)--++( -1.5,1.5);
\draw[black!60!green, line width=3pt](21,2)--++(-0.5,0.5)--++(-1,-1)--++(-1,+1)--++(-1,-1)--++(-1,+1)--++(-1,-1)--++(-1,+1)--++(-1,-1)--++(-1,+1)--++(-1,-1)--++(-2,+2)-- ++(1,1)--++(-1,1)-- ++(1,1)--++(-1,1)--++(1,1)--++(-1,1)--++(0.5,0.5);
\draw[black!60!green, line width=3pt](21,4)--++(-0.5,0.5)--++(-1,-1)--++(-1,+1)--++(-1,-1)--++(-1,+1)--++(-1,-1)--++(-1,+1)--++(-1,-1)--++(-2,+2)-- ++(1,1)--++(-1,1)--++(1,1)--++(-1,1)--++(0.5,0.5);
\draw[black!60!green, line width=3pt](21,6)--++(-0.5,0.5)--++(-1,-1)--++(-1,+1)--++(-1,-1)--++(-1,+1)--++(-1,-1)--++(-2,+2)--++(1,1)--++(-1,1)--++(0.5,0.5);
\draw[black!60!green, line width=3pt](21,8)--++(-0.5,0.5)--++(-1,-1)--++(-1,+1)--++(-1,-1)--++(-2,+2)--++(0.5,0.5);
\draw[black!60!green, line width=3pt](20,10)--++(-0.5,-0.5)--++(-0.5,0.5);
\draw[dashed, blue](0,8)--(8,0);
\draw[black, line width=4pt](8,0)--(18,10);
\end{tikzpicture}}
\caption{Solitons. The black line is either doubly occupied by a red and a green particle, or empty with no particle. It separates at region of vacuum (A) to its right from a region of vacuum (B) on its left. The other boundary of the vacuum (B) is the blue dashed line. To the left of the blue dashed line one finds again vacuum (A). The vacuum (B) forms within the light cone of the initial position of the intersection of the black two-particle or zero-particle line. }\label{Fig.5}
\end{figure}

Let us add a particle to the half filled vacuum of type A. If the initial state of the vacuum at $t_{\textup{in}}$ is green, with all sites occupied by a single green particle, the additional particle can only be red. At the position of the red particle we therefore have a two-particle state which propagates freely to the left or to the right, according to its position. We have depicted this two-particle state in Fig.~\ref{Fig.5} by a black line. The two-particle state disturbs the vacuum, as visible in Fig.~\ref{Fig.5}. While to the right of the black line one finds the vacuum A, we observe to its left the vacuum B. The vacuum B is found within the whole forward light cone of the initial position of the two-particle state. Its left boundary is indicated in Fig.~\ref{Fig.5} by a dashed blue line. (The dashed blue line is not a particle line). To the left of the dashed blue line one finds again vacuum A. The same picture obtains if instead of adding a red particle at the initial position where the black and blue dashed line meet, we take a green particle away. This is a hole excitation of the half filled vacuum A. The black line is now a line of empty positions. 

The black and dashed blue lines can be viewed as solitons separating different types of vacuum. They propagate with light velocity, similar to the fermions. They are topologically stable since matching two different types of vacuum needs a default. Solitonic excitations are the basis of Coleman's bosonization of the Thirring model~\citep{COL}. One can continue Fig.~\ref{Fig.5} to the past. The crossing of the black and blue dashed line appears as "soliton-soliton scattering".

\bigskip
\noindent
{\bf Probabilistic initial conditions and wave function}
\medskip

Arbitrary probabilistic initial conditions for the cellular automata are given by the wave function $q(t_{\textup{in}})=\big{\lbrace} q_{\tau}(t_{\textup{in}})\big{\rbrace}$. The evolution of the wave function for larger $t>t_{\textup{in}}$ obeys eq.~\eqref{eq: A02}, and for all $t$ the probabilities for finding a configuration of local occupation numbers $\tau$ is given by eq.~\eqref{eq: A05}. The expectation values of observables constructed from occupation numbers at $t$ can be obtained from the local probabilities $\big{\lbrace} p_{\tau}(t)\big{\rbrace}=\big{\lbrace} q^{2}_{\tau}(t)\big{\rbrace}$ in the standard way. This extends to time ordered products of local observables, for a discussion see ref~\citep{CWIT}.

The same wave function $q(t)$ also describes initial conditions for the fermionic quantum theory as well as their evolution to later times. The initial wave function $q(t_{\textup{in}})$ can be implemented in the fermionic quantum field theory by adding an appropriate boundary term~\citep{CWFGI} in the functional integral~\eqref{eq: 01}. The evolution of $q(t)$ is the same as for the cellular automaton, since the step evolution operator $\widehat{S}(t)$ in eq.~\eqref{eq: A02} is the same. Quantum observables constructed from local occupation numbers are represented by diagonal operators $\widehat{A}_{\tau\rho}=A_{\tau}\delta_{\tau\rho}$, with $A_{\tau}$ the value of the observable in the state $\tau$~\citep{CWIT}. Indeed, if $A$ is a function of the occupation numbers $n_{\gamma}$ the value $A_{\tau}$ obtains by inserting for $n_{\gamma}$ the value $(n_{\gamma})_{\tau}$ for the configuration $\tau$. The expectation value obeys the quantum rule
\begin{equation}\label{eq: 35}
\bigl{\langle} A(t)\bigl{\rangle}=\langle q\mid\widehat{A}\mid q \rangle=q_{\tau}\widehat{A}_{\tau\rho}q_{\rho}=A_{\tau}p_{\tau}\;.
\end{equation}
One can employ all the usual formalism of quantum mechanics, with extension to correlations for time ordered observables and additional observables represented by off-diagonal operators~\citep{CWQF}.

\bigskip
\noindent 
{\bf Schrödinger equation}
\medskip

For a large number of sites $t$ and $x$, and wave functions depending sufficiently smoothly on $x$, we can employ the continuum limit. In this limit $q(t)$ is considered as a differentiable function of time - and space - coordinates, such that lattice derivatives become partial derivatives. We can classify the states according to the conserved particle number. For one-particle states the wave function depends on the position $x$ of the particle, $q_{1}(t)=q(t,x)$\;. For two-particle state two positions have t o be specified, $q_{2}(t)=q_{2}(t,x, y)$\;, and so on. In addition, the wave functions carry "internal indices" differentiating color and left/right-movers. These indices are not displayed here. We define for $t$ even the time derivative 
\begin{equation}\label{eq: 36}
\partial_{t}q(t)=\frac{1}{4\varepsilon}\bigl(q(t+2\varepsilon)-q(t-2\varepsilon)\bigr)=W(t)q(t) \;,
\end{equation}
where eq.~\eqref{eq: A02} implies for the matrix $W$
\begin{equation}\label{eq: 37}
W(t)=\frac{1}{4\varepsilon}\bigl(\widehat{S}(t+\varepsilon)\widehat{S}(t)-\widehat{S}^{-1}(t-2\varepsilon)\widehat{S}^{-1}(t-\varepsilon)\bigr)\;.
\end{equation}
This definition takes into account that the step evolution operators differ between even and odd $t$. For $\widehat{S}(t+2\varepsilon)=\widehat{S}(t)\, , \, \widehat{S}(t+\varepsilon)=\widehat{S}(t-\varepsilon)$ and $\widehat{S}(t)=\widehat{S}^{\scriptstyle{T}}(t)$ for all $t$, the matrix $W(t) $ is antisymmetric
\begin{equation}\label{eq: 38}
W(t)=\frac{1}{4\varepsilon}\Big{\lbrace}\widehat{S}(t+\varepsilon)\widehat{S}(t)-\bigl(\widehat{S}(t+\varepsilon)\widehat{S}(t)\bigr)^{\!\scriptstyle{T}}\Big{\rbrace} \;.
\end{equation}

Eq.~\eqref{eq: 36} is the Schrödinger equation in a real formulation. (Every complex Schrödinger equation can be written as a real Schrödinger equation with twice the number of components of the wave function). For antisymmetric $W$ eq.~\eqref{eq: 36} describes a rotation of the wave function that preserves its norm. Cellular automata always produce a "unitary time evolution" in this sense.

In the presence of a complex structure the Schrödinger equation can be written as a complex differential equation, and $W$ is transformed to a hermitean Hamilton operator $H$~\citep{CWPT,CWIT}. The evolution is unitary. For the example of an one-particle state the real wave function has four components, $q(t,x)=\bigl(q_{1{\scriptscriptstyle R}}(x), q_{1{\scriptscriptstyle I}}(x),q_{2{\scriptscriptstyle R}}(x),q_{2{\scriptscriptstyle I}}(x)\bigr)$, according to the four types of particles - two colors for right movers and two colors for left movers. Combining them into a complex doublet wave function
\begin{equation}\label{eq: 39}
\psi(t, x)=\begin{pmatrix}
q_{1{\scriptscriptstyle R}}(t,x)+i \, q_{1{\scriptscriptstyle I}}(t,x) \\q_{2{\scriptscriptstyle R}}(t,x)+i \, q_{2{\scriptscriptstyle I}}(t,x)
\end{pmatrix}\;,
\end{equation}
the Schrödinger equation involves the momentum operator~$\widehat{P}$ according to
\begin{equation}\label{eq:40}
i\partial_{t}\psi=H\psi\, , \quad H=\widehat{P}\tau_{3}\,, \quad \widehat{P}=-i\partial_{x} \;.
\end{equation}
This is the free motion of massless particles - scattering plays no role for one-particle states. The interaction term will matter for the evolution of two-particle states, or states with more than two particles. We observe that the one-particle state is defined here for the totally empty vacuum. One particle excitations of other vacua will involve a more complex Schrödinger equation.

\bigskip
\noindent
\section{Generalized Ising model}\label{sec: 06}
\medskip

One more way to represent cellular automata are generalized Ising models~\citep{CWIT}. For generalized Ising models the weight function
\begin{equation}\label{eq: 41}
w[s]=\exp\big{\lbrace}-S[s]\big{\rbrace}\,,
\end{equation}
involves an action that depends on Isings spins $s_{\gamma}(t)\, , \,s^{2}_{\gamma}=1$, or associated occupation numbers $n_{\gamma}(t)=\bigl(s_{\gamma}(t)+1\bigr)/2$ ,
that take values one or zero. For generalized Ising models the weight distribution $w[s]$ is positive, $w[s]\geqslant 0$. With a proper normalization this defines a probability distribution. The generalized Ising models are standard classical statistical systems. We will construct an action $S[s]$ for which the generalized Ising model is equivalent to the cellular automaton and therefore also to the Thirring-type interacting fermionic quantum field theory.

\bigskip
\noindent 
{\bf Classical wave function and step evolution operator}
\medskip

Similarly to eqs~\eqref{eq: 15},~\eqref{eq: 16} for the fermionic model, we assume that $w[s]$ or $w[n]$ can be written as a product of local factors, with $\tilde{\mathcal{K}}(t)$ replaced by $\mathcal{K}(t)$. The local factors depend on the occupation numbers $n_{\gamma}(t), n_{\gamma}(t+\varepsilon)$ for two neighboring time layers.

We can define a "classical" wave function
\begin{equation}\label{eq: 42}
q(t)=q_{\tau}(t) h_{\tau}\bigr[n_{\gamma}(t)\bigr]=q_{\tau}h_{\tau}(t)\;,
\end{equation}
with basis functions $h_{\tau}(t)$ that are functions of the local occupation numbers $n_{\gamma}(t)$. These basis functions are constructed in close analogy to the Grassmann basis functions as a product of factors $a_{\gamma}$ for each $\gamma$. For a particle at $\gamma$ $(n_{\gamma}=1)$ one has $a_{\gamma}=n_{\gamma}$\,, while in the absence of a particle $(n_{\gamma}=0)$ the factor is $(1-n_{\gamma})$.
In comparison to the fermionic formulation, $\psi_{\gamma}$ corresponds to $1-n_{\gamma}$, and a factor $1$ at the place $\gamma$ corresponds to $n_{\gamma}$. There is no issue of signs since the integers $n_{\gamma}$ commute, and we choose the overall sign positive, e.g. $h_{\tau}\geq 0$.

For the local factors we use a double expansion, 
\begin{equation}\label{eq: 43}
\mathcal{K}(t)=h_{\tau}(t+\varepsilon)\widehat{S}_{\tau\rho}(t)h_{\rho}(t) \;.
\end{equation}
"Integrating" the occupation numbers at $t$ one has~\citep{CWIT}
\begin{equation}\label{eq: 44}
\int\! \mathcal{D}n(t)\mathcal{K}(t)q(t)=q_{\tau}(t+\varepsilon)h_{\tau}(t+\varepsilon)\;,
\end{equation}
with $\int\! \mathcal{D}n(t)=\prod_{\gamma}\sum_{n_{\gamma}(t)=0,1}\;, $
and
\begin{equation}\label{eq: 45}
q_{\tau}(t+\varepsilon)=\widehat{S}_{\tau\rho}(t)q_{\rho}(t) \;.
\end{equation}
One also finds the product rule
\begin{equation}\label{eq: 46}
\int\! \mathcal{D}n(t+\varepsilon)\mathcal{K}(t+\varepsilon)\mathcal{K}(t)= h_{\tau}(t+2\varepsilon)\widehat{S}_{\tau\rho}(t+\varepsilon)\widehat{S}_{\rho\sigma}(t)h_{\sigma}(t)\;,
\end{equation}
such that the integrated multiplication of local factors results in matrix multiplication of the step evolution operator.

For an orthogonal step evolution operator $\widehat{S}_{\tau\rho}(t)$ identical to a given fermionic model or cellular automaton, the evolution of the wave function in the generalized Ising model is the same as for the other two formulations. Also the meaning of the wave function is the same, with local probabilities $p_{\tau}(t)=q_{\tau}^{2}(t)$. For positive matrix elements $\widehat{S}_{\tau\rho}\geq 0$ we can write $\mathcal{K}(t)$ in an exponential form
\begin{equation}\label{eq: 47}
\mathcal{K}(t)=\exp\big{\lbrace} -\mathcal{L}(t)\big{\rbrace}= \exp\big{\lbrace} -h_{\tau}(t+\varepsilon)L_{\tau\rho}(t)h_{\rho}(t)\big{\rbrace}\,.
\end{equation}
The properties of the basis function imply~\citep{CWIT} for all elements of $\widehat{S}$ the relation
\begin{equation}\label{eq: 48}
\widehat{S}_{\tau\rho}(t)=\exp\big{\lbrace} -L_{\tau\rho}(t)\big{\rbrace}\;.
\end{equation}
The weight function can therefore be written in the standard form of a generalized Ising model 
\begin{equation}\label{eq: 49}
w=\exp\lbrace-S\rbrace=\exp\big{\lbrace}-\sum_{t}\mathcal{L}(t)\big{\rbrace}\;.
\end{equation}
According to eq.~\eqref{eq: 47}, $\mathcal{L}(t)$ contains interactions between occupation numbers or Ising spins on two neighboring $t$-layers.

\bigskip
\noindent 
{\bf Generalized Ising model for the Thirring type model}
\medskip

The generalized Ising models corresponding to cellular automata are constrained Ising models in the sense that many combinations of neighboring configurations $\big{\lbrace} n_{\gamma}(t)\big{\rbrace}$ and $\big{\lbrace} n_{\gamma}(t+\varepsilon)\big{\rbrace}$ are forbidden. This concerns the zero elements in the step evolution operator~$\widehat{S}_{\tau\rho}$ .  For $\widehat{S}_{\tau\rho}=0$ one takes $L_{\tau\rho}\!\rightarrow\!\infty$ , such that the probability of a configuration $\tau$ at $t + \varepsilon$ coexisting with a configuration $\rho$ at $t$ vanishes. For the allowed sequences of the cellular automaton one has $\widehat{S}_{\tau\rho}=1$ , realized by $L_{\tau\rho}=0$. In the absence of interactions the generalized Ising model for free Dirac spins~\citep{CWFGI} is given by
\begin{equation}\label{eq: 50}
\begin{split}
\mathcal{L}_{\textup{free}}(t)=&\,\beta\sum_{x}\sum_{\alpha}\Big{\lbrace} 4-n_{1\alpha}(t+\varepsilon,x+\varepsilon)n_{1\alpha}(t,x)\\
&+n_{2\alpha}(t+\varepsilon,x-\varepsilon)n_{2\alpha}(t,x)\\
&+\bigl(1-n_{1\alpha}(t+\varepsilon,x+\varepsilon)(1-n_{1\alpha}(t,x)\bigr) \\
&+\bigl(1-n_{2\alpha}(t+\varepsilon,x-\varepsilon)\bigr)\bigl(1-n_{2\alpha}(t,x)\bigr)\Big{\rbrace}\\
=&-\beta\sum_{x}\sum_{\alpha}\big{\lbrace} s_{1\alpha}(t+\varepsilon,x+\varepsilon)s_{1\alpha}(t,x)\\
&+s_{2\alpha}(t+\varepsilon,x-\varepsilon)s_{2\alpha}(t,x)-4\big{\rbrace} \;,
\end {split}
\end{equation}
taking the limit $\beta\rightarrow\infty$. Only for the allowed neighboring spin configurations $\mathcal{L}(t)$ equals zero, otherwise it diverges.

For the interaction in our model one needs to add an interaction term $\mathcal{L}_{\textup{int}}(t)$ that vanishes for the configurations of allowed color changes, and diverges for the propagation on light cones of states with single neighboring right and left movers. The latter would be allowed for free Dirac fermions, but is forbidden for our particular Thirring model. For the construction of $\mathcal{L}_{\textup{int}}(t)$  we use projectors on single particle states
\begin{equation}\label{eq: 51}
P_{k}=\dfrac{1}{2}(1-s_{k\scriptstyle{R}}s_{k\scriptstyle{I}})=n_{k\scriptstyle{R}}(1-n_{k\scriptstyle{I}})+n_{k\scriptstyle{I}}(1-n_{k\scriptstyle{R}})\;.
\end{equation}
This projector only equals one if the Ising spins $s_{k,\alpha}$ of two different colors $R$ and $I$ are opposite, which happens if either a red or a green particle is present. It vanishes if no particle or two particles of type $k$ are present. We write
\begin{equation}\label{eq: 52}
\mathcal{L}_{\textup{int}}=8\beta\sum_{x} P_{1}'P_{2}'\, L^{(2)}\, P_{1}P_{2}\;,
\end{equation}
where $P$ refers to right movers or left movers at $t$, and $P'$ correspondingly at $t+\varepsilon$. 
For $L^{(2)}$ we take
\begin{equation}\label{eq: 53}
\begin{split}
L^{(2)}=&(n_{1\scriptstyle{R}}'n_{2\scriptstyle{I}}'-n_{1\scriptstyle{I}}'n_{2\scriptstyle{R}}')(n_{1\scriptstyle{R}}n_{2\scriptstyle{I}}-n_{1\scriptstyle{I}}n_{2\scriptstyle{R}}) \\
&+(n_{1\scriptstyle{R}}'n_{2\scriptstyle{R}}'-n_{1\scriptstyle{I}}'n_{2\scriptstyle{I}}')(n_{1\scriptstyle{R}}n_{2\scriptstyle{R}}-n_{1\scriptstyle{I}}n_{2\scriptstyle{I}})\;,
\end{split}
\end{equation}
where we denote occupation numbers at $t+\varepsilon$ by $n_{k\alpha}'$ . In the presence of the projectors only one out of the four combinations $n_{1\scriptstyle{R}}n_{2\scriptstyle{R}}$, $n_{1\scriptstyle{R}}n_{2\scriptstyle{I}}$, $n_{1\scriptstyle{I}}n_{2\scriptstyle{R}}$, $n_{1\scriptstyle{I}}n_{2\scriptstyle{I}}$ can be equal to one at $t$, and similarly at $t+\varepsilon$.
If at $t+\varepsilon$ colors are exchanged, $L^{(2)}$ equals $-8\beta$. On the other hand, if the colors at $t+\varepsilon$ are the same as at $t$ the value $L^{(2)}=8\beta$ suppresses now these configuration pairs that were allowed by $\mathcal{L}_{\textup{free}}$ . For the color exchange configurations $\mathcal{L}_{\textup{free}}$ takes the value $8\beta$. This is cancelled by $\mathcal{L}_{\textup{int}}$ such that these configuration pairs are now allowed. The projections in $\mathcal{L}_{\textup{int}}$ replace effectively in $L^{(2)}$ each factor $n_{1\scriptstyle{R}}$ by $n_{1\scriptstyle{R}}(1-n_{1\scriptstyle{I}})$ etc, such that for $n_{1\scriptstyle{R}}=1$ only $n_{1\scriptstyle{I}}=0$ contributes.

Combining $\mathcal{L}(t)=\mathcal{L}_{\textup{free}}(t)+\mathcal{L}_{\textup{int}}(t)$ defines the generalized Ising model~\eqref{eq: 49}. The action involves Ising spins at two neighboring $t$ and two neighboring $x$. It can again be written as a sum over blocks. For even $t$ the blocks involve the four sites $(t,x)$, $(t, x+\varepsilon)$, $(t+\varepsilon,x)$, $(t+\varepsilon,t+\varepsilon)$ with $s_{1\alpha}=s_{\alpha}(t,x), s_{2\alpha}=s_{\alpha}(t,x+\varepsilon), s_{1\alpha}'=s_{\alpha}(t+\varepsilon,x+\varepsilon), s_{2\alpha}'=s_{\alpha}(t+\varepsilon,x)$, and similar for the occupation numbers. For odd $t+\varepsilon$ the blocks comprise the four sites $(t+\varepsilon,x), (t+\varepsilon,x-\varepsilon), (t+2\varepsilon,x), (t+2\varepsilon,x-\varepsilon)$, with the identifications $s_{1\alpha}=s_{\alpha}(t+\varepsilon,x-\varepsilon), s_{2\alpha}=s(t+\varepsilon,x), s_{1\alpha}'=s (t+2\varepsilon,x-\varepsilon), s_{2\alpha}'=s(t+2\varepsilon,x)$. We can define a coarse grained lattice with even $t$ and $x$, and combine two blocks at $(t-\varepsilon)$ and $t$, which comprise the seven sites $(t-\varepsilon, x-\varepsilon),(t-\varepsilon, x), (t, x-\varepsilon), (t,x), (t, x+\varepsilon),(t+\varepsilon,x),(t+\varepsilon,x+\varepsilon)$. On this level a distinction between even and odd $t$- layers is no longer needed. Two neighboring blocks share precisely one common site. For the example of blocks at $(t, x)$ and $(t, x+2\varepsilon) $ the spins $s_{\alpha}(t, x+\varepsilon)$ appear in both blocks. 

Combining eqs.~\eqref{eq: 50},\eqref{eq: 52} and taking $\beta\rightarrow\infty$ the generalized Ising model produces indeed the same step evolution operator as the cellular automaton and the fermionic model. The equivalence between the "functional integral" for Ising spins~\eqref{eq: 49} and the Grassmann functional integral~\eqref{eq: A10},~\eqref{eq: 23} is a first example of a fermion-bit map for a model with interactions.

The generalized Ising model is a well defined euclidean statistical model. The factor $e^{-S}$ defines a probability distribution. Many powerful methods of statistical mechanics and condensed matter physics can be applied in this setting, as block spinning or (functional) renormalization. In particular, one can perform Monte Carlo simulations, for example for large finite $\beta$ and take the limit $\beta\rightarrow\infty$.

\bigskip
\noindent
\section{Discussion}\label{sec: 07}
\medskip

We have proposed a simple two-dimensional model for which three equivalent formulations exist: 
\begin{enumerate}
\item a probabilistic classical cellular automaton, 
\item a Grassmann functional for fermions, 
\item a generalized Ising model.
\end{enumerate}
This allows for a simultaneous use of methods from all three fields. We emphasize that the generalized Ising model is a standard euclidean functional integral formulated on a square lattice. On the other hand, the Grassmann functional exhibits Lorentz symmetry in the continuum limit, with its characteristic Minkowski signature differentiating between time and space. Also the cellular automaton singles out time as the direction in which the steps of the automaton are performed. Our example shows that there is no contradiction between these concepts.

Our model of interacting fermions is an example how quantum mechanics emerges from classical statistics ~\citep{CWEM,CWQMCS}. A quantum field theory is quantum mechanics for many particles. The quantum mechanics of a single particle is a special case. The evolution of the one-particle wave function obeys a standard Schrödinger equation, and similarly for two-particle states and so on.  All features of quantum mechanics as non-commuting operators and the quantum laws for the computation of expectation values emerge in a natural way~\cite{CWPT,CWFCS,CWIT}. There are no contradictions to no-go theorems~\citep{CWEM,CWQMCS}. Even though our model is simple, it exhibits non-trivial features of interacting quantum field theories as spontaneous symmetry breaking or solitons. The one-particle Schrödinger equation depends on the ground state. 

Our construction of a cellular automaton may appear somewhat special. For the same cellular automaton one can use a different construction of alternating step evolution operators for the propagation and the interaction, that will be published elsewhere. We believe that a wide range of fermionic quantum field theories can be constructed as cellular automata in this way.


\bibliography{probabilistic_cellular_automata}

\end{document}